\documentclass[12pt]{article}

\usepackage{amsmath}
\allowdisplaybreaks[4]

\usepackage{amssymb}
\usepackage{epsfig}

\begin{document}

\newcommand{\half}{\frac{1}{2}}

%%%%%%%%%%%% Equation numbering:          
\renewcommand{\theequation}{\thesection.\arabic{equation}}
%%%%%%%%%%%%

\newcommand{\ope}{OPE}
\newcommand{\hwr}{highest weight representation}
\newcommand{\hws}{highest weight state}
\newcommand{\desc}{descendant}

\newcommand{\f}{\phi}
\newcommand{\p}{\psi}

\newcommand{\dd}{\partial}
\newcommand{\Vb}{V_{\vec \beta}}

\newcommand{\barray}{\setlength\arraycolsep{2pt} \begin{array}}
\newcommand{\earray}{\end{array}}
\newcommand{\dis}{\displaystyle}

\newcommand{\ab}[2]{ \big|\, #1 \, , #2 \, \big>}
\newcommand{\ket}[1]{ \big|\, #1  \, \big>}
\newcommand{\abi}[2]{ \left\{ #1 \, , #2 \right\}}

\newcommand{\Rab}[4]{
\left| 
\setlength\arraycolsep{0.2ex}
\begin{array}{ccc}
#1&,& #2\\[0.4ex]
#3&,& #4\\  
\earray
\right>}

\newcommand{\NO}[2]{{:} #1 #2 {:} }
\newcommand{\NOthree}[3]{{:} #1 #2 #3 {:}}
\newcommand{\com}[2]{\left[#1,#2\right]}
\newcommand{\acom}[2]{\{#1,#2\}}

\newcommand\sw{${\mathcal{SW}}(3/2,2)$\ }  
\newcommand\cl{ c_p^{(1)} }
\newcommand\cu{ c_p^{(2)} }

\newcommand{\aup}{a_\uparrow}
\newcommand{\adown}{a_\downarrow}
\newcommand{\bup}{b_\uparrow}
\newcommand{\bdown}{b_\downarrow}
\newcommand{\avec}[1]{{\vec \alpha}_{#1}}

\newcommand{\ii}{\mathrm{i}}
\newcommand{\ee}{\mathrm{e}}

%%%%%%%%%%%%%%%%%%%%%%%%%%%%%%%%%%%
%%%%%%%%%%%%%%%%%%%%%%%%%%%%%%%%%%%

\title{
Unitary representations of $\mathcal{SW}(3/2,2)$\\  superconformal algebra}
\author{
Doron Gepner,
Boris Noyvert\\ \\[-2ex]
{\it \small Department of Particle Physics,}\\
{\it \small Weizmann Institute of Science,} \\
{\it \small 76100, Rehovot,  Israel.}
}
\date{}

\maketitle

\begin{abstract}
$\mathcal{SW}(3/2,2)$ superconformal algebra is $\mathcal{W}$ algebra
with two Virasoro ope\-rators. The Kac determinant is calculated and the 
complete list of unitary representations is determined.
Two types of extensions of $\mathcal{SW}(3/2,2)$ algebra are
discussed. 
A new approach to construction of $\mathcal{W}$
algebras from rational conformal field theories
is proposed.

\end{abstract}

%\tableofcontents

\section{Introduction}

\setcounter{equation}{0}

After the fundamental work of Zamolodchikov
\cite{Zamolodchikov:1985wn}
conformal $\mathcal{W}$ algebras become the subject of great interest 
in many branches of physics and mathematics.
Many $\mathcal{W}$ algebras are known (for review see
\cite{Bouwknegt:1993wg}).
However there is no complete classification of $\mathcal{W}$ algebras,
and representation theory only of some of them is developed
\cite{Fateev:1987vh, Fateev:1988zh, Lukyanov:1990tf}.
The study of minimal models \cite{BPZ} of $\mathcal{W}$ algebras is relevant
for the classification problem of rational conformal field theories.

In this paper we study \sw extended superconformal algebra, generated by
two supercurrents of dimensions $\frac{3}{2}$ and $2$.
%% (sections \ref{sec:sw-algebra} and \ref{constr}).
%% This is the first $\mathcal W$ algebra with two Virasoro
 %% generators.
The algebra was first constructed in
Ref.~\cite{fofs}.

At central charge $c=12$ the algebra has geometrical meaning,
it is associated to 8-dimensional manifolds of $Spin(7)$ 
holonomy \cite{shv, shv2}. These manifolds are relevant for
string theory. Sigma models based on  $Spin(7)$ holonomy manifolds
can provide a superstring compactification scheme 
down to 2 dimensions, leaving minimal number of
space--time supersymmetries. In such a compactification \sw algebra
plays the same role, as $N=2$ superconformal algebra in
compactifications on Calabi--Yau manifolds.

Our main subject is the representation theory of \sw algebra. 
We define \hwr s of the algebra, calculate their Kac determinant,
study conditions of unitarity and list all the unitary
representations (Appendix \ref{apB}).

The two supercurrents contain four 
conformal fields of scaling dimensions $\frac{3}{2}$,
$2$, $2$, $\frac{5}{2}$.  Two dimension 2 generators 
can be represented as two commuting Virasoro ope\-rators.
One of these Virasoro subalgebras is in the rational regime, since its 
central charge is less than 1.
This is a crucial point in the study of unitary representation theory 
of \sw algebra. Unitarity is constrained to two series of models 
at central charges 
\begin{equation}  \label{cp}
\begin{aligned}                                      
 \cl &= 6- \dis\frac{18}{p+1} \\ 
 \cu &= 6+\dis \frac{18}{p}
\end{aligned}
 \qquad p=3,4, 5 \ldots
\end{equation}
together with their limiting point $c=6$.
There is no continuum of allowed by unitarity central charges.
(In some sense, the continuum shrinks to the point $c=6$.)

At fixed central charge from (\ref{cp}) the unitary
spectrum consists of discrete points and continuous lines
in the two dimensional space of weights.
In parti\-cu\-lar, any representation with nonnegative
scaling dimension ($h \ge 0$) is allowed.
In Section~\ref{unitar} we decompose unitary 
\hwr s of \sw algebra to representations 
of the rational Virasoro subalgebra.
The decomposition involves all the representations from one row
(at $c=\cl$) or from one column (at $c=\cu$)
of the Kac table of the Virasoro algebra minimal model.
A subset of the discrete spectrum representations
forms minimal models of \sw algebra
(Section~\ref{sec:Fusion-rules-minimal}).

In order to eliminate the continuous spectrum
one should consider extensions of \sw algebra.
Two types of extensions of \sw algebra are constructed
in Section~\ref{section:extensions}, relating \sw to two 
other extended superconformal algebras 
$\mathcal{SW}(3/2,3/2,2)$ 
\cite{Blumenhagen:1992nm} and 
$\mathcal{SW}^2(1,2)$ 
\cite{Romans:1992wi}. (See \cite{Bouwknegt:1993wg} for notations.)
The second extension leads to interesting class of models possessing
$N=2$ superconformal symmetry.

The paper is organized as follows. Section~\ref{sec:sw-algebra}
introduces the operator product expansions (\ope s) of \sw algebra.
The bosonic subalgebra is decomposed to two commuting
Virasoro algebras in Section~\ref{sec:Virasoro-algebra-emb}.

In Section~\ref{constr} we propose a new way of construction
of \sw algebra from minimal models of the Virasoro algebra.
The similar approach can be applied to construction of new 
 $\mathcal{W}$ algebras.

The \hwr s of \sw are defined in Section~\ref{sec:High-weight-repr}.
There are two sectors: Neveu--Schwarz (NS) and Ramond,
each is labeled by two weights.
Coulomb gas representation is redeveloped in 
Section~\ref{sec:Free-field-repr},
and the Kac determinant is calculated in 
Section~\ref{sec:Degen-repr-Kac}.

The restrictions of unitarity on \hwr s are analyzed
in Section~\ref{unitar}. We illustrate the calculations 
by two examples: $c=12/5$ and $c=12$ models.
Fusion rules are discussed in Section~\ref{sec:Fusion-rules-minimal}
and extensions of \sw algebra -- in Section~\ref{section:extensions}.

Most of the work is done
with a help of {\sl Mathematica} package \cite{Thielemans:1991uw}
for symbolic computation of operator product expansions.

%%%%%%%%%%%%%%%%%%%%%%%%%%%%%

\section{\sw algebra}              \label{sec:sw-algebra}

\setcounter{equation}{0}

Motivated by interplay between string theory and geometry
Shatashvili and Vafa \cite{shv,shv2} studied extended symmetry 
algebras, which underlie sigma models with $N=1$
superconformal symmetry on manifolds of exceptional
holonomy, namely 7--dimensional manifolds of 
$G_2$ holonomy and 8--dimensional manifolds of 
$Spin(7)$ holonomy.

The symmetry algebra contains Virasoro generator $T$ and supersymmetry
ge\-ne\-rator $G$. In the $Spin(7)$ holonomy case the existence of  $Spin(7)$  invariant $4$--form 
leads to dimension $2$ operator $\tilde X$ \cite{shv}. It has a superpartner $\tilde M$
of dimension $\frac{5}{2}$. These four operators constitute a closed conformal
algebra with central charge $c=12$. The subalgebra generated by $\tilde X$
coincides with Ising model, i.e. Virasoro algebra with central charge $\half$.

As pointed out in \cite{Figueroa-O'Farrill:1997hm} 
the algebra is a special ($c=12$) case of the generic algebra,
which was obtained in 
%%can be generalized to an arbitrary value 
%%of central charge. 
%The generalization is an extension of the 
%$N=1$ superconformal algebra, 
%including weight $2$ superconformal primary. 
%We will call it super $W_2$ algebra. 
%%The algebra was obtained in 
Ref.~\cite{fofs} using a conformal bootstrap 
method and was called by authors  super $W_2$ algebra.
We adopt the conventions of \cite{Bouwknegt:1993wg} and call 
it \sw algebra.

The algebra  is an extension of  
$N=1$ superconformal algebra by dimension~2 superconformal multiplet,
which consists of Virasoro primary field $W$ of scaling dimension
$\Delta_W=2$ and its superpartner $U$ of dimension
$\Delta_U=\frac{5}{2}$:
\begin{align}
G(z)\,W(w)&= \frac{U(w)}{z-w}   \, ,       \label{opeGW} \\
G(z)\,U(w)&=\frac{2\, \Delta_W \, W(w)}{(z-w)^2}+
\frac{\dd W(w)}{z-w}  \, .
\end{align}
Our normalization of $U$ differs from one in  
Ref.~\cite{fofs} by
factor $-\ii \sqrt{5}$. 
The \ope\ of $W$ with itself is 
\begin{align}
W(z) \, W(w) & = \dis{\frac{c/2}{(z-w)^4}}
%%+{ \frac{1}{(z-w)^2} \, 
+ \frac{
  2 \, T(w) +{\frac{2 \, (6 + 5\,c) }
   {{\sqrt{15 - c}}\,{\sqrt{21 + 4\,c}}}} \, W(w) }{(z-w)^2} + \nonumber\\
%%+{ \frac{1}{z-w} \, 
& \hspace{25ex} +\frac{
\dd T(w) +{\frac{6 + 5\,c}
   {{\sqrt{15 - c}}\,{\sqrt{21 + 4\,c}}}}\, \dd W(w) }{z-w}\, .
\end{align}

The important property of the algebra 
that it closes nonlinearly: the singular terms of 
\ope s
$W(z) \, U(w)$ and  $U(z) \, U(w)$ contain normal ordered 
products of basic fields. The \ope s are given in
Appendix~\ref{apA}.

The mode expansions of generators are defined in the usual way:
\begin{equation}
\mathcal{O}(z)=\sum_n \mathcal{O}_n \,   z^{-n-\Delta_{\mathcal{O}}} \, ,
\end{equation}
where $\mathcal{O}$ is any generator of the algebra, and
$\Delta_{\mathcal{O}}$ is its scaling dimension.
(The modes of $T(z)$ are denoted by $L_n$.)
Unitarity is introduced by standard conjugation relation
$\mathcal{O}_n^\dagger = \mathcal{O}_{-n}$ for any generator 
except $U$. \sw commutation relations are consistent with 
nonstandard conjugation property $U_n^\dagger = -U_{-n}$.

%%%%%%%%%%%%%%%%%%%%%%%%%%%%%

\section{Virasoro algebra embeddings}       \label{sec:Virasoro-algebra-emb}

\setcounter{equation}{0}

The algebra contains a few different subalgebras. These are Virasoro subalgebra generated
by $T(z)$, superconformal subalgebra generated by $T(z)$ and $G(z)$,
and bosonic subalgebra generated
by $T(z)$ and $W(z)$. Let's discuss the last one.

One can introduce instead of $T$ and $W$ two new dimension $2$ operators:
\begin{equation}
\begin{aligned}
T &= A+B \, ,\\
W&=  \sqrt{\tfrac{21+4 \, c}{15-c}} \, A-   \sqrt{\tfrac{15-c}{21+4 \, c}}\, B \, .
\end{aligned}
\end{equation}
As a result the bosonic part of \sw is decomposed
to two commuting Virasoro algebras $A$ and $B$ with central charges
\begin{equation}   
%\barray{rcl}
c_A=\dis{\frac{c \,(15-c)}{3 \,(12+c)}}\, , \qquad
c_B=\dis{\frac{c \,(21+4 \, c)}{3 \,(12+c)}} \, .
%\earray
\end{equation}
We will call the $A$ and $B$ Virasoro algebras the internal and external algebras. 
%The \sw algebra in terms of operators $T,A,G,U$ is presented in the Appendix~\ref{apA}.

Unitary representations of an algebra are necessarily unitary representations of all
its subalgebras. In the region $c \ge 0$ we have $0 \le c_A \le 1$. Therefore by
applying the nonunitarity theorem 
\cite{Friedan:1984xq,Friedan:1986kd}
to $A$ Virasoro subalgebra  we deduce that
there are no unitary representations of \sw out of special values of central charge
  $c_A=1$ and
\begin{equation}              \label{caunitary}
c_A (c) =1-\frac{6}{p(p+1)}\, , \qquad p=3,4,5 \ldots
\end{equation}
In the latter case (\ref{caunitary}) there is a finite number of unitary
\hwr s of internal Virasoro algebra: $\Phi_{m,n}$ \cite{BPZ}. Their weights
form Kac table:
\begin{equation}                                         \label{amn}
a_{m,n}=\frac{ {\left( m\,p - 
        n\,\left( 1 + p \right)  \right) }^2-1}
    {4\,p\,\left( 1 + p \right) }\, , \qquad m=1, \ldots \! , p \, , 
\quad n=1, \ldots\! , p-1 \, .
\end{equation}
Resolving equation (\ref{caunitary}) for $c$ we get two series of the
unitary models (\ref{cp}).
The series $\cl$ and $\cu$ start from $3/2$ and $12$  and converge to $c=6$
from below and from up respectively. The limiting point $c=6$ corresponds
to the value $c_A=1$ and also is not forbidden by the nonunitarity theorem.

Other subalgebras give no new restrictions on unitarity. Indeed
in the range $3/2 \le c \le 12$ the representations 
of superconformal algebra are unitary.
And since $c_B \ge 1$ in this range the $B$ subalgebra is also unitary.

As one could expect the $c=3/2$ model can be written in terms of one free boson and 
one free fermion:
\begin{equation}
\barray{lcl}                                                \label{c32}
A=\half  \NO{\p}{\dd \p} &\qquad &B=-\half  \NO{\dd \f}{\dd \f}\\
G=\NO{\p}{\dd \f}&\qquad & U=-\sqrt{2} \NO{\dd \p}{\dd \f}+\frac{1}{\sqrt{2}}
\NO{\p}{\dd ^2 \f} 
\earray
\end{equation}
where the \ope s of the free fields are given by
\begin{align}   
     \label{fp1}                         
\f(z) \,\f(w) &= - \log (z-w) \, ,\\
     \label{fp2}   
\p(z) \,\p(w) &= -\frac{1}{z-w} \, .
\end{align}   

In this simple example  the fermionic part of the stress-energy tensor 
is the internal Virasoro operator and its bosonic part is the external Virasoro operator.
The internal Virasoro part is identical to the Ising model.

%%%%%%%%%%%%%%%%%%%%%%%%%%%%%%%%%%

\section{Different construction of \sw algebra}                    \label{constr}

\setcounter{equation}{0}

\sw algebra was first constructed in Ref.~\cite{fofs} by conformal
bootstrap method. In this section we present another 
(probably, less formal) way of constructing the algebra.

As we have seen in the previous section, the bosonic part of \sw
splits to two commuting Virasoro algebras, $A$ and $B$. Other generators
of the algebra should fall in some representation of $A \oplus B$ algebra.
Indeed $G(z)$ is a primary field of $A \oplus B$ with dimensions 
%$\abi{\frac{15 - c}{2\,\left( 12 + c \right) }}{\frac{3}{2}-\frac{15 - c}{2\,\left( 12 + c \right) }}$
$(a_{1,2},\frac{3}{2}-a_{1,2})$ at $c=\cl$ and  
$(a_{2,1},\frac{3}{2}-a_{2,1})$ at $c=\cu$. 
%%Based on this fact,
%%we propose a different from 
%%Ref.~\cite{fofs} 
%%way of construction of \sw algebra.

The starting point of the construction is
Virasoro algebra $A$ in the minimal model regime, i.e. the value
of central charge $c_A$ is given by (\ref{caunitary}). We want to extend the algebra
to include the minimal model primary field $\Phi_{1,2}$. From other point
of view we want this field to be superconformal generator of dimension $\frac{3}{2}$.
This is possible, if A is not the full Virasoro operator, but there is another 
Virasoro algebra $B$ (commuting with $A$) which completes the full Virasoro operator
$T=A+B$.  
Now we can introduce the field $G(z)$, which is primary with respect
to $A$ with weight $a_{1,2}$ and primary with respect to $T$ with weight $\frac{3}{2}$.
The \ope s are:
\begin{align}
T(z)\, G(w)&=\frac{\frac{3}{2} \, G(w)}{(z-w)^2}+\frac{\dd G(w)}{z-w}
\, ,\\
A(z)\, G(w)&=\frac{a_{1,2} \, G(w)}{(z-w)^2}+
\frac{(A_{-1}G)(w)}{z-w}\, . 
\end{align}
$( A_{-1}G)$ is a new generator of the algebra of the full dimension $\frac{5}{2}$.
The \ope\ of $A(z)$ with $( A_{-1}G)(w)$ includes the field $( A_{-1}^2 G)(w)$,
but this is not a new generator due to existence of null vector on level 2 of $\Phi_{1,2}$
representation of $A$ Virasoro algebra:
\begin{equation}
%\left( A_{-1}^2 - \frac{13-c_A+\sqrt{(25-c_A)(1-c_A)}}{12} A_{-2} \right) \Phi_{1,2}=0
\left( A_{-1}^2 - \frac{2 +4 \, a_{1,2} }{3}\, A_{-2} \right)
\Phi_{1,2}=0 \, .
\end{equation}
So $( A_{-1}^2 G)$ is proportional to $( A_{-2} G)\,$, which can be written
as $\NO{A}{G}$. We need to introduce only one new field $U$:
\begin{equation}
U=\frac{1}{N_U} \left( (A_{-1} G)- \frac{2}{3} \, a_{1,2}\, \dd G
\right) .
\end{equation}
We add $\dd G $ term in order to make $U$ field primary with respect
to $T$, and $N_U$ is just a normalization factor.
With this definition we write the \ope s:
\begin{align}
A(z)\, G(w)&=\frac{a_{1,2} \, G(w)}{(z-w)^2}+\frac{N_U \, U(w)+ 
\frac{2}{3} \, a_{1,2}\, \dd G(w)}{z-w} \, ,\\
T(z)\, U(w)&=\frac{\frac{5}{2} \, U(w)}{(z-w)^2}+\frac{\dd
  U(w)}{z-w} \, ,
\\
%\end{align}
%\\[-6ex]
%\begin{multline}
%\phantom{\hspace{-\hoffset} A(z)\, U(w)}
A(z)\, U(w)&=\frac{\frac{2}{N_U}\, a_{1,2}\, (1-\frac{2}{3} \,
  a_{1,2}) \, G(w)}{(z-w)^3}+
\nonumber \\
&+\frac{(1+\frac{a_{1,2}}{3})\, U(w)+ 
\frac{2}{3 \,N_U} \, a_{1,2} \,(1-\frac{2}{3} \, a_{1,2}) \, \dd
G(w)}{(z-w)^2}+
\nonumber \\
& + \frac{
\frac{2}{3 \,N_U} \, (1+2\, a_{1,2}) \, \NO{A}{G} (w)
-\frac{2}{3} \, a_{1,2}\, \dd U(w)-\frac{4}{9 \,N_U} \, a_{1,2} ^2 \, \dd^2 G(w)
}
{z-w}  \, . 
\end{align}
%\end{multline}
This set of \ope s guarantees that Jacobi identities of type $[X,X,Y]$ are satisfied,
where $X$ stands for $T$ or $A$, $Y$ -- for $G$ or $U$.

Now we need to close the algebra by specifying the \ope s
of type $Y(z) Y(w)$.
%$G(z) G(w) , \, G(z)U(w) , \, U(z)U(w)$.
From the point of view of $A$ Virasoro algebra 
$G$ and $U$ are fields in $\Phi_{1,2}$ representation,
$A$ and $T$ are in $\Phi_{1,1}$  (vacuum) representation.
The fusion \cite{BPZ}
\begin{equation}
\Phi_{1,2} \times \Phi_{1,2}= \Phi_{1,1}+\Phi_{1,3}
\end{equation}
includes the new representation $\Phi_{1,3}$.
It should enter to the \ope s of type $Y(z) Y(w)$.
%$G(z) G(w) , \, G(z)U(w) , \, U(z)U(w)$.
It is easy to check that fields \mbox{$\NO{G}{\dd G}$} and $\NO{G}{U}$
are in the $\Phi_{1,3}$ representation. We
construct  $G(z) G(w)$, $G(z)U(w)$, $U(z)U(w)$ as most general 
\ope s including only $A$, $T$ operators with their derivatives and composites
and \mbox{ $\NO{G}{\dd G}$,}  $\NO{G}{U}$.
After that the arbitrary coefficients are fixed as functions of $c$ and $a_{1,2}$ 
by Jacobi identities
of type $[ X,Y,Y ]$.
Jacobi identities of type $[ Y,Y,Y ]$ also fix the
connection between the full central charge and  
$a_{1,2}$:
\begin{equation}
a_{1,2}=\frac{15 - c}{2\,\left( 12 + c \right) } \, .
\end{equation}
Finally we obtain \sw algebra in the form presented in  Appendix~\ref{apA}.

We started from discrete unitary values of $c_A$, but the algebra
can be continued smoothly to any $c_A \le 1$, i.e.
to any $c$.
Another remark: one could start from  $\Phi_{2,1}$
representation instead of $\Phi_{1,2}$,
but the same algebra is obtained.

Highest weight representations of the algebra are introduced
in the next section, but already at this stage we can predict some 
representations. There is a magic relation between
the dimensions of any Kac table (\ref{amn}):
\begin{equation}        \label{magic_relation}
 a_{1,n}+a_{m,1}-a_{m,n}=\frac{ (m-1)(n-1)}{2} \qquad \forall p \, .
\end{equation}
The right hand side of this equation is integer or half-integer.
Taking into account that the relation (\ref{magic_relation})
corresponds to the fusion rule \cite{BPZ}
\begin{equation}
   \Phi_{1,n} \times \Phi_{m,1} = \Phi_{m,n} \, ,
\end{equation}
the fields $\Phi_{1,n}$ and $\Phi_{m,1}$ are found to be local
or semilocal with respect to each other.

The \sw generator $G$ behaves as $ \Phi_{1,2}$ under internal 
Virasoro symmetry. The purely internal representations of
$A \oplus B$ algebra $(\Phi_{m,1}, 0)$ are local 
(odd $m$) or semilocal (even $m$) with respect to 
supercurrent $G$. They are first candidates to
Neveu--Schwarz (NS) and Ramond representations respectively.
Indeed we find them in the discrete unitary spectrum
(see end of Section~\ref{unitar}). Moreover this 
set of fields forms minimal models of \sw algebra 
(Section~\ref{sec:Fusion-rules-minimal}).

The type of construction, discussed in this section,
can be used for building new conformal algebras.
For example, one can take $N=1$ superconformal minimal model
as internal algebra, and to extend  it by simplest nontrivial 
NS field $\Phi_{1,3}$. We expect that
$\mathcal{SW}(3/2,3/2,2)$  conformal algebra 
(\cite{Blumenhagen:1992nm} and 
Section~\ref{section:extensions} of this paper)
can be constructed this way.
But this is a subject to another paper.

The superconformal algebras of 
Refs.~\cite{Fradkin:1992bz,Fradkin:1992km}
can also be treated as extensions of rational conformal models.
The role of internal algebra is played by affine Lie algebra $\widehat g$.
It is extended by supercurrents $G_i$ in some 
(usually fundamental, see \cite{Fradkin:1992bz,Fradkin:1992km})
representation of $g$.
The dimension of supercurrents is $\frac{3}{2}$ 
with respect to the full Virasoro generator  $T$,
which can be seen as a sum of two parts: $T=A_g+B$,
where $A_g$ is Sugawara energy-momentum tensor
of $\widehat g$ and $B$ is external Virasoro operator, 
commuting with $A_g$ and with all the currents of  $\widehat g$ .
The algebra closes nonlinearly:
the \ope s $G_i(z) \, G_j(w)$ involve normal orderings
of the $\widehat g$ currents.

%%%%%%%%%%%%%%%%%%%%%%%%%%%%%%%%%%

\section{Highest weight representations}        \label{sec:High-weight-repr}

\setcounter{equation}{0}

%There are two sectors in the theory: NS and Ramond sectors.
One can define consistently two modings of \sw algebra. In both cases 
modes of the bosonic generators are integer. Modes of the fermionic
generators can be chosen half integer (NS sector) or integer (Ramond sector).
There are no other possible modings of the algebra.  

\subsection{NS sector}

There are two zero modes: $L_0$ and $A_0$.
Since their commutator is zero, they gene\-ra\-te the Cartan subalgebra.
Following notations of \cite{shv} we label the \hwr s by
eigenvalues of $A_0$ and $B_0$:
\begin{equation}
\begin{aligned}
A_0 \ab{a}{b} &= a \ab{a}{b} \, ,\\
B_0 \ab{a}{b} &= b \ab{a}{b} \, ,\\
{\mathcal{O}}_n \ab{a}{b}&=0 \qquad \forall n > 0\, ,
\end{aligned}
\end{equation}
and $\mathcal{O}$ being any operator of the algebra. $a$ is called internal dimension
and $b$ is called external dimension of the state. Also we will often use
the full scaling dimension $h=a+b$, the eigenvalue of $L_0$.

The primary field $\Psi(z)$ corresponding to the \hws\ $\ab{a}{b}$ has the following 
\ope s with generators of the algebra:
\begin{equation}  \label{opeNS}
\begin{aligned}
T(z) \, \Psi(w)&=\frac{h \, \Psi(w)}{(z-w)^2}+\frac{\dd
  \Psi(w)}{z-w}+\ldots \\
A(z) \, \Psi(w)&=\frac{a \, \Psi(w)}{(z-w)^2}+\frac{(A_{-1} \Psi)(w)}{z-w}+\ldots\\
G(z) \, \Psi(w)&=\frac{(G_{-1/2} \Psi)(w)}{z-w}+\ldots\\
U(z) \, \Psi(w)&=\frac{(U_{-1/2} \Psi)(w)}{(z-w)^2}+
\frac{(U_{-3/2} \Psi)(w)}{z-w}+\ldots
\end{aligned}
\end{equation}

%The necessary condition for one dimensional representation to be unitary
%is that its internal dimension $a$ is included in the Kac table of dimensions of correspondent
%Virasoro unitary minimal model:
%\begin{equation}                                         \label{amn}
%a_{m,n}=\frac{ {\left( m\,p - 
%        n\,\left( 1 + p \right)  \right) }^2-1}
%    {4\,p\,\left( 1 + p \right) } \qquad m=1 \ldots p \quad n=1\ldots  p-1
%\end{equation}
%And of course the external dimension $b=h-a$ should not be negative. 

At this point we are ready to write down the NS representations of the $c=3/2$ model (\ref{c32}):
\begin{equation}
\barray{lcl}                                 \label{c32NS}
\ab{0}{0}&\qquad &\rm{vacuum}\\[0.6ex]
 \ab{\half}{0} &\qquad &\p(z)\\[0.6ex]
\ab{\frac{1}{16}}{\frac{1}{16}}&\qquad &\sigma(z)\, s(z)\\[0.6ex]
\ab{0}{\half \beta^2}&\qquad &\ee ^{\ii \beta \f(z)}
\earray
\end{equation}
where $\sigma(z)$ and $s(z)$ are twist fields of free fermion and free boson theory
respectively.  They are defined by \ope s (see e.g., \cite{Ginsparg:1988ui}):
\begin{align}
\label{twist_sigma}
\p(z) \, \sigma(w) &\sim \frac{\mu(w)}{(z-w)^{1/2}}+\ldots \\
\label{twist_s}
\dd \f (z) \, s(w) &\sim \frac{\tau(w)}{(z-w)^{1/2}}+\ldots 
\end{align}
The dimensions of $\sigma , \mu , s , \tau$ are $\frac{1}{16}, \frac{1}{16},\frac{1}{16},\frac{9}{16}$
respectively.

The representations (\ref{c32NS}) are obviously unitary by construction.
The unitary spectrum contains continuous line:
$\ab{0}{b \ge 0} $ and two discrete points: $\ab{\half}{0}$
and $ \ab{\frac{1}{16}}{\frac{1}{16}}$. As we will see all other unitary models 
also consist of discrete states and continuum.

\subsection{Ramond sector}

The \hws\ is annihilated by all the operators with positive indices. The zero mode 
algebra here is more complicated than in the NS sector.
There are four zero modes:
$L_0, A_0, G_0, U_0 \, $.  $L_0$ is commutative with three other zero modes,
so it can be represented by a number $h$. The commutation relations between 
$A_0, G_0, U_0$ are nonlinear, they contain normal ordered products
$ \NO{A}{G}_0 \,$, $\NO{T}{A}_0 \, $, $\NO{G}{U}_0 \, $, $\NO{G}{\dd G}_0 \,$,
which are infinite sums in terms of modes of generators (see appendix~\ref{apC} for exact 
definitions). 
However, we always keep in mind that we apply
these operators to \hws s. Any positive mode acting on \hws\ 
gives zero.
So in the \hwr\  only a few terms from the infinite sums will survive
and the (anti)commutators of the zero modes will be:
\begin{equation}                        \label{zma}
\begin{aligned}
\com{A_0}{G_0}&=
-\frac{ {\sqrt{15 - c}}\,
      {\sqrt{21 + 4\,c}}  }
    {3\,\left( 12 + c \right) }\,{U_0} \, ,\\ 
\com{A_0}{U_0}&=
  \frac{3\,{\sqrt{15 - c} }\,
     \left( 3 + c \right) \,{G_0}}{4\,
     \left( 12 + c \right) \,
     {\sqrt{21 + 4\,c}}} + 
  \frac{9\,{U_0}}{12 + c}
-\frac{54 \, G_0 \, A_0 }{{\sqrt{15 - c}}\,
     {\sqrt{21 + 4\,c}}} \, ,\\ 
\acom{G_0}{G_0}
&= -\frac{c}{12} + 2\, h  \, ,\\ 
\acom{G_0}{U_0}&=0 \, ,\\
\acom{U_0}{U_0}
&=-\frac{3\,\left( 3 + c \right) \,
     \left( c - 24\,h \right) }{16\,
     \left( 21 + 4\,c \right) } \\
&\quad + 
  \frac{54\,{G_0}\,{U_0}}{{\sqrt{15 - c}}\,
     {\sqrt{21 + 4\,c}}} +
  \frac{27\,\left( 12 + c \right) \,
     \left( c - 24\,h \right) \,{A_0}}{2\,
     \left( 15 - c \right) \,
     \left( 21 + 4\,c \right) } \, .
\end{aligned}
\end{equation}
They define finite dimensional quadratic $\mathcal{W}$ superalgebra.

The irreducible representations of this algebra are two dimensional.
We choose $A_0$ to be diagonal:
\begin{equation}
A_0=\left(
\barray{cc}
a&0\\
0&\, \, a+\Delta a\\
\earray
\right) ,
\end{equation}
where $\Delta a$ is some function of $a,h,c$.
Then $G_0$ and $U_0$ are:
\begin{equation}
G_0=\left(
\barray{cc}
0&g\\
g&\, 0\\
\earray
\right) ,
\qquad
U_0=\left(
\barray{cc}
0&u\\
-u&\, 0\\
\earray
\right) .
\end{equation}
$(U_0^\dagger=-U_0)$. The commutation relations (\ref{zma})
are satisfied by
\begin{equation}
g=\sqrt{h-c/24} \quad \text{and} \quad u=g \, \Delta a \, 
\frac{3\,\left( 12 + c \right) }
  {{\sqrt{15 - c}}\,{\sqrt{21 + 4\,c}}}\, .
\end{equation}
There are two solutions for $\Delta a$:
\begin{equation}                                                        \label{deltaa}
\Delta a_{1,2}= \frac{9 \pm {\sqrt{{\left( 6 - c \right) }^2 + 
        72\,a\,\left( 12 + c \right) }}}{2\,
    \left( 12 + c \right) } \, .
\end{equation}
Note that $\Delta a$ doesn't depend on $h$, it depends only on $a$ and $c$.

Ramond sector is well defined only for $h\ge c/24$. In the special case $h= c/24$
the representation of the zero mode algebra (\ref{zma}) can be chosen
one dimensional with
\begin{equation}
G_0=U_0=0, \qquad A_0=a\, .
\end{equation}
We will call such representation Ramond ground state by analogy
to $N=2$ superconformal algebra.

Acting by negative modes of generators on the representation of the zero mode algebra
we get the full tower of states of Ramond representation. We will denote Ramond 
ground \hws s by $\ab{a}{b} \, (a+b=c/24)$ and two dimensional Ramond
\hws s
by $\Rab{\aup}{\bup}{\adown}{\bdown}$, 
where $\aup$ and $\adown$
are two eigenvalues of $A_0$, $\bup$ and $\bdown$
are two eigenvalues of $B_0$ ($\aup + \bup =\adown +\bdown=h$).
$\aup$ and $\adown$ are not independent,
their difference is fixed by $\adown - \aup = \Delta a_{1,2}$ from (\ref{deltaa}).

The \ope s of generators of \sw with primary field of Ramond type are:
\begin{equation}
\begin{aligned}
T(z) \, \Psi(w)&=\frac{h \, \Psi(w)}{(z-w)^2}+\frac{\dd \Psi(w)}{z-w}+\ldots\\
A(z) \, \Psi(w)&=\frac{(A_0 \Psi)(w)}{(z-w)^2}+\frac{(A_{-1} \Psi)(w)}{z-w}+\ldots\\
G(z) \, \Psi(w)&=\frac{(G_{0} \Psi)(w)}{(z-w)^{3/2}}+
\frac{(G_{-1} \Psi)(w)}{(z-w)^{1/2}}+\ldots\\
U(z) \, \Psi(w)&=\frac{(U_{0} \Psi)(w)}{(z-w)^{5/2}}+
\frac{(U_{-1} \Psi)(w)}{(z-w)^{3/2}}+\ldots
\end{aligned}
\end{equation}

%Ramond representation can be unitary only if both $A_0$ eigenvalues $\aup$
%and $\adown$ belong to the Kac table of correspondent Virasoro unitary minimal
%model (\ref{amn}) and if both $B_0$ eigenvalues $h-\aup$
%and $h-\adown$ are positive or zero. 

Coming back to the $c=3/2$ example (\ref{c32}) 
we write unitary Ramond representations of the model:
\begin{equation}                              \label{c32R}
\barray{lcl}                            
\Rab{\half}{\frac{1}{16}}{0}{\frac{9}{16}}&\qquad &
\left( \barray{c} \p(z) \, s(z) \\[0.4ex] \tau(z)\\  \earray \right) \\[2ex]
\ab{0}{\frac{1}{16}} &\qquad & s(z) \\[0.6ex]
\ab{\frac{1}{16}}{0} &\qquad & \sigma(z)\\[0.6ex]
\Rab{\frac{1}{16}}{\half \beta^2}{\frac{1}{16}}{\half \beta^2}& \qquad &
\left( \barray{c} \sigma(z)\\[0.4ex] \mu(z) \earray \right) \ee^{\ii \beta \f(z)}\\
\earray
\end{equation}
with twist fields defined in (\ref{twist_sigma}) and  (\ref{twist_s}).

%%%%%%%%%%%%%%%%%%%%%%%%%%%%%%%%%%%%%%%%%%%%%%%%%

\section{Free field representation}      \label{sec:Free-field-repr}

\setcounter{equation}{0}

The free field representation of \sw algebra was derived by several
authors \cite{Komata:1991cb,Mallwitz:1995hh}.
They get the covariant representation, i.e. in terms of two 
$N=1$ free superfields: $\f + \theta \, \p$. We don't introduce here the
superspace formalism but use explicit noncovariant formulation.

The construction of $N=1$ superalgebra in terms of free bosons and fermions
is well known:\footnote
{Usually the basis in the space of bosons is chosen in such a way,
that there is no $\dd^2 \f_2$ term in the expression for $T$. 
%Our choice just leads to simplification of the formulas below.
We choose rotated basis for later convenience.}  
\begin{equation}
\begin{split}
T &= \half  \,\NO{\p_1}{\dd \p_1}+\half  \,\NO{\p_2}{\dd \p_2}\\
&-\half  \,\NO{\dd \f_1}{\dd \f_1} -\half \,\NO{\dd \f_2}{\dd \f_2}
-\frac{\ii \left( 6 - c \right) }
   {6\,{\sqrt{3 + c}}}\, \dd^2 \f_1 - 
  \frac{3 - 2\,c}{6\,{\sqrt{3 + c}}}\, \dd^2 \f_2 \, ,\\
G &= \NO{ \p_1}{\dd \f_1} +\NO{ \p_2}{\dd \f_2} +\frac{\ii \left( 6 - c \right) }
   {3\,{\sqrt{3 + c}}}\, \dd \p_1 + 
  \frac{3 - 2\,c}{3\,{\sqrt{3 + c}}}\, \dd \p_2 \, ,
\end{split}
\end{equation}
where the free fields are defined by (\ref{fp1}) and (\ref{fp2}).

We look for $W$ being a linear combination 
(with $c$--dependent coefficients)
of all possible dimension $2$ composite fields constructed from 
the free fields.
$W$ must satisfy the \ope s with $T$, $G$ and itself, from these conditions
we calculate the coefficients. 
The result is:
\begin{equation}
\barray{rcl}
W&=&\frac{1}{{\sqrt{15 - c}}\, {\sqrt{21 + 4\,c}}} \, \bigg(
\left( 3 - 2\,c \right) \NO {\dd \f_1} {\dd \f_1}
+ \, \half \, \left( 15 - c \right) \NO {\dd \f_2} {\dd \f_2}\\[0.6ex]
&&-9\, \sqrt{3 + c} \, \NOthree {\psi_ 1}{\psi_ 2}{\dd \f_1}-
\left( 3 - 2\,c \right) \NO {\psi_ 1} {\dd \psi_ 1}+
\left( 15 - c \right)  \NO {\psi_ 2} {\dd \psi_ 2} \\[0.6ex]
&&+ 3\, \ii \,\left( 6 - c \right)  \NO {\psi_ 2} {\dd \psi_ 1}+
\frac{\ii \, \left( 6 - c \right) \, \left( 3 - 2\,c \right)}{3 \, \sqrt{3 + c}} \, \dd^2 \f_1+
\frac{\left( 15 - c \right) \, \left( 3 - 2\,c \right)}{6 \, 
\sqrt{3 + c}}\, \dd^2 \f_2 \bigg)\, .
\earray
\end{equation}
The free field representation of $U$ is obtained automatically
from \ope\ (\ref{opeGW}).

Primary fields in NS sector are represented by free field exponentials:
  \begin{equation}
\Vb (z)={:}\exp\left( \ii {\vec \beta}\! \cdot \! {\vec \f}(z)
\right){:}  \, ,\qquad 
{\vec \beta}\! \cdot \! {\vec \f }= \beta_1 \, \f_1 + \beta_2 \, \f_2
\, .
\end{equation}
Applying $T(z)$ and $A(z)$ to the exponentials we get the weights parameterized by
\begin{align} 
                                                \label{hs1s2NS}
h({\vec \beta}) &=\half \, \beta_1
\left(\beta_1+\frac{6 - c}{3\,{\sqrt{3 + c}}} \right)+
\half \, \beta_2  \left(\beta_2 - 
\ii \, \frac{3 - 2\,c}{3\,{\sqrt{3 +c}}} \right) ,\\%[0.97ex]
                                                 \label{as1s2NS}
a({\vec \beta}) &=\frac{3 + c}{2\,\left( 12 + c \right) } \, \beta_1 
\left(\beta_1+\frac{6 - c}{3\,{\sqrt{3 + c}}} \right) .
\end{align}
In Ramond sector there are also fermionic zero modes. We can choose
two Pauli matrices $\sigma_x$ and $\sigma_y$ to represent $\p_1$
and $\p_2$ zero modes. $A_0$ and $L_0$ are diagonal in such a 
representation. So the Ramond states are constructed from
two dimensional spin space and free field exponentials:
\begin{equation}
\left(
\barray{c}
 \ket{\uparrow } \\[0.6ex]    \ket{\downarrow }
\earray
\right)
 \Vb (z) \, .
\end{equation}
Both states have the same full dimension:
 \begin{equation}                      \label{hs1s2R}
h({\vec \beta}) =\half \, \beta_1
\left(\beta_1+\frac{6 - c}{3\,{\sqrt{3 + c}}} \right)+
\half \, \beta_2  \left(\beta_2 - \ii \, \frac{3 - 2\,c}{3\,{\sqrt{3 + c}}} \right) +\frac{1}{8} \, ,
\end{equation}
which differs from the expression in NS sector by $\frac{1}{8}$ ($\frac{1}{16}$
from each fermion).
The internal dimensions of $\ket{\uparrow } $ and $\ket{\downarrow } $ states
are
%\begin{equation}                    \label{as1s2R}
%\barray{rcl}
%a({\vec \beta}, \uparrow )&=& \frac{3 + c}{2\,\left( 12 + c \right) } 
%\left(\beta_1+\frac{15 - c}{6\,{\sqrt{3 + c}}} \right)^2 - 
%\frac{\left( 6 - c \right)^2}{72 \, \left( 12 + c \right)} \, ,\\[0.97ex]
%a({\vec \beta}, \downarrow )&=& \frac{3 + c}{2\,\left( 12 + c \right) }
%\left(\beta_1 - \frac{\sqrt{3 + c}}{6} \right)^2 - 
%\frac{\left( 6 - c \right)^2}{72 \, \left( 12 + c \right)} \, .
%\earray
%\end{equation}
\begin{equation}                    \label{as1s2R}
\begin{split}
a({\vec \beta}, \uparrow )&= \frac{3 + c}{2\,\left( 12 + c \right) } 
\left(\beta_1+\frac{15 - c}{6\,{\sqrt{3 + c}}} \right)^2 - 
\frac{\left( 6 - c \right)^2}{72 \, \left( 12 + c \right)} \, ,\\[0.97ex]
a({\vec \beta}, \downarrow )&= \frac{3 + c}{2\,\left( 12 + c \right) }
\left(\beta_1 - \frac{\sqrt{3 + c}}{6} \right)^2 - 
\frac{\left( 6 - c \right)^2}{72 \, \left( 12 + c \right)} \, .
\end{split}
\end{equation}

Note that in both sectors
the expressions for dimensions have two independent  $\mathbb{Z}_2$ 
symmetries:\footnote{In Ramond sector the first symmetry interchanges 
$\ket{\uparrow } $ and $\ket{\downarrow } $ states.}
\begin{align}
                    \label{sym1}
\beta_1 & \to -\frac{6 - c}{3\,{\sqrt{3 + c}}}- \beta_1  \, ,\\
                    \label{sym2}
\beta_2 & \to  \ii \, \frac{3 - 2\,c}{3\,{\sqrt{3 + c}}}-\beta_2 \, .
\end{align}
One can identify 4 representations, related by the  $\mathbb{Z}_2 \times \mathbb{Z}_2$
symmetry.

%%%%%%%%%%%%%%%%%%%%%%%%%%%%%%%%%%%%

\section{Degenerate representations and Kac\\ determinant} 

\label{sec:Degen-repr-Kac}

\setcounter{equation}{0}

Degenerate representations of various conformal algebras were 
described in 
\cite{Feigin:1982st,Bershadsky:1985dq,Mussardo:1987eq,
Fateev:1987vh,
Fateev:1988zh,Lukyanov:1990tf,
Mussardo:1989av}
using the technique of Coulomb gas formalism.
We will follow the same methods.

Screening operators $Q$ are operators which commute with all the algebra:
 \begin{equation}                                 \label{scr}
\com{Q}{{\mathcal{O}}_n} = 0 \quad \forall \, n \, .
\end{equation}
The form of $Q$
\begin{equation}
Q_{{\vec \alpha}}=\oint \! \mathrm{d} z \, \Theta_{\vec \alpha}(z) \, ,
\qquad
\Theta_{{\vec \alpha}}(z)= \,
\NO{{\vec \alpha}\cdot {\vec \p}(z) \, }{{ \ee^{\ii \, {\vec \alpha}\cdot {\vec \f}(z)}}}  \, ,
\end{equation}
where $h({\vec \alpha})= {\half}$,
ensures the zero commutation relations with $L_n$ and
$G_n$. The condition $\com{Q}{W_n}= 0$ fixes the vector $\vec \alpha$. 
There are $3$ screening operators: 
$Q_{\avec{1}}, \, Q_{\avec{2}}, \, Q_{\avec{3}}, $ where
\begin{equation}                                                             \label{alpha}
\barray{rcl}
\avec{1}&=& \frac{3}{{\sqrt{3 + c}}}\, \Big( 1 \, , \, - \ii \Big)  ,\\[1ex]
\avec{2}&=& \frac{3}{{\sqrt{3 + c}}}\, \Big( -1 \, , \, 0 \Big)  ,\\[1ex]
\avec{3}&=& \frac{{\sqrt{3 + c}}}{3}\, \Big(1\, , \, 0 \Big)  .
\earray
\end{equation}

The first two vectors (with omitted $c$-dependent coefficients) can be identified with
simple roots of $osp(3|2) \approx B(1,1)$ Lie superalgebra, quantum
Drinfeld--Sokolov reduction of which is exactly \sw conformal algebra
\cite{Komata:1991cb}.

One can construct null states acting by screening operators on primary field:
\begin{equation}                                                                                                              \label{eta}
\barray{rcl}
\eta_n &=
&\left( {Q_{\vec \alpha}} \right)^n \, 
V_{{\vec \beta}- n {\vec \alpha}} (0) = \\
&=& \dis{\oint_{C_1}} \! \mathrm{d} z_1 \, \Theta_{\vec \alpha}(z_1) \cdots 
\dis{\oint_{C_n}} \! \mathrm{d} z_n  \,  
\Theta_{\vec \alpha}(z_n) \, V_{{\vec \beta}- n {\vec \alpha}} (0) \, .
\earray 
\end{equation}
The state is a \desc\ of $\Vb$, but it behaves as $V_{{\vec \beta}- n {\vec \alpha}}$
under action of generators of the algebra:
\begin{equation}
\barray{rcl}
{\mathcal{O}}_k \, \ket{\eta_n}  &=& 0 \, , \quad k>0 \, ,\\
L_0 \, \ket{\eta_n} &=& h({\vec \beta}- n \, {\vec \alpha}) \, \ket{\eta_n} \, ,\\
A_0 \,  \ket{\eta_n} &=& a({\vec \beta}- n \, {\vec \alpha}) \, \ket{\eta_n} \, .\\
\earray
\end{equation}
This means that the representation constructed from $\Vb$
is reducible and $\ket{\eta_n}$ must be a null vector
%$|\, \eta_n \! >$ is a null vector in the space constructed from $\Vb$,
on the level 
\begin{equation}                                                                                               \label{N}
N= h({\vec \beta}- n \, {\vec \alpha})-h({\vec \beta}) \, .
\end{equation}

The definition (\ref{eta}) contains $n$-multiple integral over the contour
on some Riemann surface \cite{Dotsenko:1984nm}. 
Such a nontrivial closed contour exists and the integration is
well defined, if the following condition is satisfied:
\begin{equation}                                                                          \label{betas}
{\vec \beta} \cdot {\vec \alpha} = 
- \frac{1+m}{2} + \frac{1+n}{2} \, {\vec \alpha \, }^2 \, ,
\quad m \, , n \in \mathbb{Z} \, .
\end{equation}
The level (\ref{N}), in which the null state appears is 
\begin{equation}
\begin{aligned}
N&= { \frac{n \, m}{2}} \qquad ({\vec \alpha \,}^2 \ne 0) \, ,\\
N&= {\frac{|m|}{2}} \qquad ({\vec \alpha \,}^2 =0) \, .\\
\end{aligned}
\end{equation}

If there is a null state on the level $N$, the Kac determinant on this level
vanishes. The vanishing curves of the Kac determinant are given
in the table.
%Table~\ref{Kacs1s2}.
%\begin{table}[h]                                
%$$ 
\begin{equation}        \label{Kacs1s2}           
\begin{array}{|l |c |c |c |}
\hline
0=&N \mathrm{(level)}&\mathrm{NS}&\mathrm{Ramond}\\
\hline
&&&\\[-2.3ex]
\beta_1-\ii \, \beta_2 + \frac{\sqrt{3 + c}}{3} \, \frac{\left( 1 -l  \right)}{2}&
\frac{|l|}{2}&l \in 2 \mathbb{Z}+1&l \in 2 \mathbb{Z}\\[0.6ex]
\beta_1- \frac{\sqrt{3 + c}}{3}\, \frac{\left( 1 - m  \right)}{2}
+\frac{3}{\sqrt{3 + c}}\, \frac{\left( 1 - n  \right)}{2}&
{\frac{n \, m}{2}} >0&m+n \in 2 \mathbb{Z}& m+n \in 2 \mathbb{Z}+1\\[0.6ex]
\ii \, \beta_2 - \frac{\sqrt{3 + c}}{3} \, \left( 1 - j  \right) + 
\frac{3}{\sqrt{3 + c}}\, \frac{\left( 1 +k  \right)}{2}
& j \, k >0 &j, k \in \mathbb{Z}& j, k \in \mathbb{Z}\\[0.6ex]
\beta_1+ \ii \, \beta_2 - 
\frac{\sqrt{3 + c}}{3} \, \frac{\left( 3 -l  \right)}{2} + \frac{3}{\sqrt{3 + c}}&
\frac{|l|}{2}&l \in 2 \mathbb{Z}+1&l \in 2 \mathbb{Z}\\[0.8ex]
\hline
\earray
\end{equation}
%$$
%\caption{Vanishing curves of  the Kac determinant of \sw %in terms of $\beta_1, \beta_2$     
%algebra.}\label{Kacs1s2} 
%\end{table}
The first two vanishing curves are obtained by inserting 
$\avec{1}$ and $\avec{2}$ in (\ref{betas}). The third 
screening vector $\avec{3}$ gives the same equation as $\avec{2}$.
As we have already mentioned $\avec{1}$ and $\avec{2}$
are simple roots of $B(1,1)$, $\avec{3}$ is a negative root,
a pair of $\avec{2}$.
$B(1,1)$ Lie superalgebra has 5 positive roots, 2 even: 
$\avec{2}, 2 \, (\avec{1}+\avec{2})$ and 3 odd:
$\avec{1},  \avec{1}+\avec{2},  \avec{1}+2\, \avec{2}$.
The last two lines in (\ref{Kacs1s2}) correspond to positive roots 
$\avec{1}+\avec{2}$ and $\avec{1}+2\, \avec{2}$
respectively. 
They can be obtained by inserting $Q = Q_{\avec{1}}\, Q_{\avec{2}}$
and $Q = Q_{\avec{1}}\, Q_{\avec{2}}^2$ in (\ref{eta}).
The negative roots give the same expressions as their positive 
counterparts. 
The fifth positive root $2 \, (\avec{1}+\avec{2})$ corresponds
to composite screening operator $Q = \left( Q_{\avec{1}}\, Q_{\avec{2}} \right)^2$,
which produces the same null states as $Q = Q_{\avec{1}}\, Q_{\avec{2}}$.

The last vanishing curve can be also obtained from the
first one by applying the 
$\mathbb{Z}_2$ symmetry transformation (\ref{sym1}). Moreover the full set of 
4 vanishing curves is invariant under $\mathbb{Z}_2 \times \mathbb{Z}_2$
symmetry (\ref{sym1}, \ref{sym2}).

Using (\ref{hs1s2NS}),  (\ref{as1s2NS}), (\ref{hs1s2R}) and (\ref{as1s2R})
one can easily express
the Kac determinant in $a, h$ parameterization. As a result
we get three types of vanishing curves in the 3-dimensional space
of $c, a, h$.

\noindent {\bf NS sector}:
\begin{equation}                     \label{kacbeg}
%$$
\barray{|rcl|}
\hline
&&\\[-2.7ex]
f^{\mathrm{NS}}_{m, n}&=& a + \frac{1}{18\,\left( 12 + c \right) }
  \left( \left( 3 + c \right) \, \frac{1 - m }{2} - 9\,\frac{1 - n }{2} \right)
  \left( \left(3 + c \right) \frac{ 1 + m }{2} - 9\,\frac{1 + n  }{2} \right)\\
\multicolumn{3}{|c|}{m, n>0 \qquad m+n \in 2 \mathbb{Z} \qquad
  N=\frac{m\, n}{2}}\\[0.6ex]
\hline
&&\\[-2.7ex]
g^{\mathrm{NS}}_{j, k}&=& a \, (12 + c)-(1 + h)\,(3 + c) +\\
&&\frac{1}{18}\left( \left( 3 + c \right) \, \left(1 - j \right) - 9\,\frac{1 - k }{2} \right)
  \left( \left( 3 + c \right) \, \left(1 + j \right) - 9\,\frac{1 + k }{2} \right)\\
\multicolumn{3}{|c|}{j, k>0  \qquad N=j \, k}\\[0.6ex]
\hline
&&\\[-2.7ex]
d^{\mathrm{NS}}_{l}&=&h^2 - \frac{a}{18}  (12 + c)\, l^2+
\frac{h}{12}  \left( 3-c+\frac{3+c}{3} \, l^2 \right)+\\
&&\frac{1}{576}\left( 1 - l^2  \right)
  \left( (3-c)^2 -\frac{(3 + c)^2}{9}\, l^2 \right)\\
\multicolumn{3}{|c|}{l >0  \qquad l \in 2 \mathbb{Z}+1 \qquad
  N=\frac{l}{2}}\\
\hline
\earray
%$$
%\caption{Vanishing curves of  the Kac determinant of 
%NS sector of \sw algebra in terms of $a, b$.}
\end{equation}

\noindent{\bf Ramond sector}:
\begin{equation}
%$$
\barray{|rcl|}
\hline
&&\\[-2.7ex]
f^{\mathrm{R}}_{m, n}&=&a+ \frac{(6-c)^2}{72 \, (12+c)}+
\frac{3}{\sqrt{2\,(12+c)}} \, r\,\sqrt{a+\frac{(6-c)^2}{72 \, (12+c)}}+\\
&&\frac{1}{72 \, (12+c)} \left( 9\,(1+n)-(3+c)\, m \right) 
\left(9\,(1-n)+(3+c)\, m \right)\\
\multicolumn{3}{|c|}{m, n>0 \qquad m+n \in 2 \mathbb{Z}+1 \qquad
  N=\frac{m\, n}{2}}\\[0.6ex]
\hline
&&\\[-2.7ex]
g^{\mathrm{R}}_{j, k}&=&a \,(12 + c) - h \,(3 + c) +
\frac{3 \sqrt{12 + c}}{\sqrt{2}} \, r \, \sqrt{a+\frac{(6-c)^2}{72 \, (12+c)}}-\frac{12 + 7 c}{8}+\\
&&\frac{1}{72} \left(2 \, (3 + c) (1-j) + 9 \, (1-k)\right)
\left(2 \, (3 + c) (1+j) + 9 \, (1+k)\right)\\
\multicolumn{3}{|c|}{j, k>0  \qquad N=j \, k}\\[0.6ex]
\hline
&&\\[-2.7ex]
d^{\mathrm{R}}_{l}&=&h^2 -  \frac{\sqrt{12 + c}}{6\sqrt{2}}\, l^2 \, r\,
\sqrt{a+\frac{(6-c)^2}{72 \, (12+c)}}-\\
&&\frac{h}{36}\left(3 \, c - (3 + c)\, l^2\right)-
\frac{a}{18}(12 + c)l^2 +\\[0.2ex]
&&\frac{1}{5184}\left(9 \, c^2 -2 \, (234 - 15 \, c+5 \, c^2 )l^2+ (3+c)^2 l^4 \right)\\
\multicolumn{3}{|c|}{l > 0  \qquad l \in 2 \mathbb{Z} \qquad N=\frac{l}{2}}\\[0.6ex]
d^{\mathrm{R}}_{0}&=&h-c/24\\
\hline
\earray
%$$
\end{equation}
where $r= \pm 1$ distinguish between 
$\ket{\uparrow } $ and $\ket{\downarrow } $ states
of Ramond sector.

The Kac determinant on level $N$ is (up to $c$--dependent
factor)
\begin{equation}
\barray{c}
\dis
\mathrm{det} M^{\mathrm{NS}}_N (c,a,h)=
\prod_{\scriptstyle 
1 \le m \, n\le 2\, N 
\atop \scriptstyle 
m+n \, \mathrm{even}
}(f^{\mathrm{NS}}_{m, n})^{P_\mathrm{NS} (N-\frac{m\,n}{2})} \times 
\\ \dis 
\prod_{1 \le j\,k \le N}
(g^{\mathrm{NS}}_{j, k})^{P_\mathrm{NS} (N-j \, k)}
\prod_{\scriptstyle 
1 \le l \le 2\, N 
\atop \scriptstyle 
l \, \mathrm{odd}
}(d^{\mathrm{NS}}_{l})^{{\widetilde P}_\mathrm{NS} (N-\frac{l}{2}, l)}
\earray
\end{equation}
for NS sector. The counting of states is given by generating partition functions:
\begin{equation}
\barray{rcl}
\dis \sum_n P_\mathrm{NS} (n) \, x^n &=& \dis
\prod_{k=1}^\infty \left( \frac{1+x^{k-1/2}}{1-x^k} \right)^2 \, ,\\
\dis \sum_n {\widetilde P}_\mathrm{NS} (n, l) \, x^n &=& \dis
\frac{1}{1+x^{l/2}} \prod_{k=1}^\infty \left(
  \frac{1+x^{k-1/2}}{1-x^k} \right)^2 \, .
\earray
\end{equation}

The Kac determinant of Ramond type representation:
\begin{equation}
\barray{c}
\dis
\mathrm{det} M^{\mathrm{R}}_N (c,a,h)=
\prod_{\scriptstyle 
1 \le m \, n\le 2\, N 
\atop \scriptstyle 
m+n \, \mathrm{odd}
}(f^{\mathrm{R}}_{m, n})^{P_\mathrm{R} (N-\frac{m\,n}{2})} \times 
\\ \dis 
\prod_{1 \le j\,k \le N}
(g^{\mathrm{R}}_{j, k})^{P_\mathrm{R} (N-j \, k)}
\prod_{\scriptstyle 
0 \le l \le 2\, N 
\atop \scriptstyle 
l \, \mathrm{even}
}(d^{\mathrm{R}}_{l})^{{\widetilde P}_\mathrm{R} (N-\frac{l}{2}, l)}
\, ,
\earray
\end{equation}
where the generating functions are:
\begin{equation}                   \label{kacend}
\barray{rcl}
\dis \sum_n P_\mathrm{R} (n) \, x^n &=& \dis
\prod_{k=1}^\infty \left( \frac{1+x^{k}}{1-x^k} \right)^2  ,\\
\dis \sum_n {\widetilde P}_\mathrm{R} (n, l) \, x^n &=& \dis
\frac{1}{1+x^{l/2}} \prod_{k=1}^\infty \left( \frac{1+x^{k}}{1-x^k}
\right)^2  .
\earray
\end{equation}

Computer calculations of the Kac determinant on the first few levels
verify the formulae  (\ref{kacbeg}--\ref{kacend}).

%%%%%%%%%%%%%%%%%%%%%%%%%%%%%%%%%%%%%

\section{Unitary representations}                    \label{unitar}

\setcounter{equation}{0}

Now we are ready to discuss the unitarity restrictions on NS and Ramond representations.
Unitary representations appear only at discrete set of values of central charge (\ref{cp}).
The necessary condition for representation to be unitary
is that its internal dimension $a$ is included in the Kac table of dimensions of correspondent
Virasoro unitary minimal model (\ref{amn}).
And of course the external dimension $b$ should not be negative. 

Two dimensional Ramond representation can be unitary only if both $A_0$ eigenvalues $\aup$
and $\adown=\aup+ \Delta a$ belong to the Kac table 
 (\ref{amn}) and if both $B_0$ eigenvalues $\bup=h-\aup$
and $\bdown=h-\adown$ are positive or zero. 

The natural question arises: what value of $a+ \Delta a$
do we get choosing $a$ from the Kac table? It is simple exercise to check that
\begin{equation}                 \label{shiftR1}
a_{m,n}+\Delta a(a_{m,n}, \cl) = a_{m,n \pm 1}\, ,
\end{equation}
where $\pm$ corresponds to two choices of $ \Delta a$ in (\ref{deltaa}).
Analogously for the descending series:
\begin{equation}                 \label{shiftR2}
a_{m,n}+\Delta a(a_{m,n}, \cu) = a_{m \pm 1,n }\, .
\end{equation}
So, we see that the pair of Ramond internal dimensions in the $\cl$ ($\cu$)
model is a pair of horizontal (vertical) neighboring dimensions from the
correspondent Kac table.

Since $G$ and $U$ currents are in $\Phi_{1,2}$ %($\Phi_{2,1}$)
representation of the Virasoro minimal model (Section~\ref{constr}),
the property (\ref{shiftR1}) is explained by 
the minimal model fusions of $\Phi_{1,2}$:
\begin{equation}                                          \label{shift-fusion}
  \Phi_{1,2} \times \Phi_{m,n} = \Phi_{m,n-1} + \Phi_{m,n+1} \, .
\end{equation}

We illustrate it on two examples, one from ascending, one from descending
series of models.

\begin{description}

\item[Example 1. ] $\quad c=12/5$

The simplest model is $c=3/2$ model (\ref{c32}). It is explicitly solved by free 
field representation (\ref{c32NS}, \ref{c32R}). So we take as example the next model $c=12/5$.
Its internal Virasoro part is the Tricritical Ising model ($c_A=7/10$), the Kac table of which is:
\begin{equation}
\left(
\begin{array}{ccc}
  0 & \frac{7}{16} & \frac{3}{2} \\[0.97ex]
\frac{1}{10} & \frac{3}{80} & \frac{3}{5} \\[0.97ex]
\frac{3}{5} & \frac{3}{80} & \frac{1}{10}  \\[0.97ex]
\frac{3}{2} & \frac{7}{16} & 0 
\end{array}
\right) .
\end{equation}
In the NS sector the representation $\ab{a}{b}$ can be unitary only
for $b \ge 0$ and $a$ from the set 
\begin{equation}
0 \quad \frac{3}{80} \quad \frac{1}{10} \quad \frac{7}{16} \quad
\frac{3}{5} \quad \frac{3}{2} \, .
\end{equation}
Ramond ground states ($a+b=\frac{1}{10}$):
\begin{equation}
\ab{\frac{1}{10}}{0}\qquad \ab{\frac{3}{80}}{\frac{1}{16}} \qquad
\ab{0}{\frac{1}{10}}  \, .
\end{equation}
For two dimensional Ramond states the pairs of internal dimensions 
$\Big\{ \dis {\aup \atop \adown} \Big\}$ are:
\begin{equation}
\Big\{ \dis  {0 \atop \frac{7}{16}} \Big\} \quad
\Big\{ \dis { \frac{7}{16} \atop \frac{3}{2}} \Big\} \quad
\Big\{ \dis { \frac{1}{10} \atop \frac{3}{80}} \Big\} \quad
\Big\{ \dis { \frac{3}{80} \atop \frac{3}{5}} \Big\} \, .
\end{equation}

\item[Example 2.] $\quad c=12$

The simplest example from descending series is $c=12$ model.
Its internal subalgebra is the Ising model ($c_A=1/2$). This example is 
exactly the case of $Spin(7)$ algebra discussed in \cite{shv}. The Kac table
of the Ising model:
\begin{equation} \left(
\begin{array}{cc} 
0 & \frac{1}{2} \\[0.97ex]  
\frac{1}{16} & \frac{1}{16} \\[0.97ex]
 \frac{1}{2} & 0
\end{array}
\right).
\end{equation}
So, the unitary internal dimensions are 
\begin{equation}
0 \quad \frac{1}{16} \quad \frac{1}{2} \, .
\end{equation}
Ramond ground states ($h=c/24=1/2$):
\begin{equation}
\ab{\half}{0} \qquad \ab{\frac{1}{16}}{\frac{7}{16}} \qquad
\ab{0}{\half} \, .
\end{equation}
The pairs of Ramond $A_0$ eigenvalues:
\begin{equation}
\Big\{   {0 \atop \frac{1}{16}} \Big\} \quad
\Big\{  { \frac{1}{16} \atop \half} \Big\} \, .
\end{equation}

\end{description}

The \hwr\ of \sw can be decomposed to a direct sum of 
representations of its internal subalgebra. The weights of all representations
in the decomposition should also belong to the unitary set of dimensions (\ref{amn}).
We have to study the \desc s of \hws\ of \sw algebra, which are 
\hwr s with respect to internal subalgebra $A$. The simplest examples 
of such states are the level  $\half$ \desc s of NS \hws. They are annihilated
by positive $A$ modes:
\begin{equation}
A_n \, G_{-\half} \ab{a}{b}=A_n \, U_{-\half} \ab{a}{b}=0 \qquad \forall n >0\, .
\end{equation}
However these states are not $A_0$ eigenstates. But one can construct two
linear combinations of these  states, which  are $A_0$ eigenstates
and therefore \hwr s of  the internal Virasoro algebra:
\begin{equation}
\barray{c}
\ket{\chi_{1,2}}=\left(\frac{12+c \, \pm \, 3\, {\sqrt{{\left( 6 - c \right) }^2 + 
        72\,a\,\left( 12 + c \right) }}}
{2\,\sqrt{15-c}\,\sqrt{21+4 \, c}} \,
G_{-1/2}+U_{-1/2} \right) \ab{a}{b}\, , \\
A_0 \, \ket{\chi_{1,2}}=
{\widetilde a}_{1,2} \, \ket{\chi_{1,2}}\, ,
\earray
\end{equation}
where $\widetilde a$ is a function of $a,b,c$. 
Providing that $a$ is one of the Kac dimensions
$\widetilde a$ has also to be included in the Kac table. 
$B_0$ eigenvalue ${\widetilde b}=h+\half-{\widetilde a}$ should be nonnegative.
$\Delta a_{1,2}={\widetilde a}_{1,2}-a$
is the same as the difference of Ramond internal dimensions (\ref{deltaa}).

For the case of interest, $c=\cl$ ($c=\cu$)
the $\ab{a_{m,n}}{b} \,$ state has two level $\half$ \desc s, which
are $A$ primaries with internal dimensions being
nearest to $a_{m,n}$ neighbors in the Kac table: ${\widetilde a}=a_{m,n +1}$ 
and ${\widetilde a}=a_{m,n-1}$ 
($a_{m+1,n}$ and $a_{m-1,n}$), 
where we define $a_{m,0}= a_{m+p,p-1}$ using periodicity of Kac
dimensions.
Again the shift in the Kac table is explained by fusion (\ref{shift-fusion}).

We have to take into account another possibility that one of the states
$\ket{\chi_{1,2}}$ is null and should be modded out to make the 
representation irreducible. In such a case there is only one \desc\ on level $\half$.
It happens when the values of $a,b,c$ lie on the zero locus of the Kac determinant
on level $\half$: %det$M^{\mathrm{NS}}_{1/2}=0$.
$f_{1,1}=0$ (it's essentially $a=0$ for every $c$) and $d_1=0$.

At $c=\cl$ ($c=\cu$) the states $\ab{a_{m,1}}{b}$ and $\ab{a_{m,p-1}}{b}$ 
($\ab{a_{1,n}}{b}$ and $\ab{a_{p,n}}{b}$)
from the first and last columns (rows)
of the correspondent Kac table generically are not unitary, since they have one of the 
\desc s, that is ``neighbor'' from outside of the Kac table. ``Generically'' is in
the meaning that this is not true for intersections with $f_{1,1}=0$ and $d_1=0$,
where the nonunitary \desc\ 
becomes null. The hall line $\{ a=a_{1,1}=0, \,
b \ge 0 \}$ should not be considered as nonunitary, because it coincides with the line
$f_{1,1}=0$.

\begin{description}
 
\item[Example 1.] $\quad c=12/5$

All the region below $d_1=0$ curve (Figure~\ref{fig125}) is nonunitary,
since The Kac determinant on level $\half$ is negative.
$\ab{\frac{1}{10}}{b}$ has level $\half$ \desc s with $A_0$ weights 
${\widetilde a}_1 =a_{2,0}=a_{3,4}=\frac{63}{80}$, 
${\widetilde a}_2 =a_{2,2}=\frac{3}{80}$;
the \desc s of $\ab{\frac{3}{5}}{b}$:
${\widetilde a}_1 =a_{3,0}=a_{2,4}=\frac{143}{80}$, 
${\widetilde a}_2 =a_{3,2}=\frac{3}{80}$;
the \desc s of $\ab{\frac{3}{2}}{b}$:
${\widetilde a}_1 =a_{4,0}=a_{1,4}=\frac{51}{16}$, 
${\widetilde a}_2 =a_{1,2}=\frac{7}{16}$.
In all three cases ${\widetilde a}_1$ is not from the unitary set,
therefore $\ab{\frac{1}{10}}{b}$, $\ab{\frac{3}{5}}{b}$ and
$\ab{\frac{3}{2}}{b}$ are not unitary except the case then
$\ket{\chi_1}$ is null. It happens only on $d_1=0$ curve:
$\ab{\frac{1}{10}}{\frac{1}{10}}$ and $\ab{\frac{3}{5}}{0}$.
$a=\frac{3}{2}$ doesn't intersect $d_1=0$, therefore $\ab{\frac{3}{2}}{b}$
are all nonunitary.
The nonunitary \desc\ of $\ab{0}{b}$ is null for every $b$, the state stays unitary.

States with weights $a=\frac{7}{16}$ and  $a=\frac{3}{80}$
are from the middle of the Kac table, $\ab{\frac{7}{16}}{b}$  has \desc s
on level $\half$ with ${\widetilde a}_1 = 0$ and  ${\widetilde a}_2 = \frac{3}{2}$,
$\ab{\frac{3}{80}}{b}$ has \desc s with  
${\widetilde a}_1 = \frac{1}{10}$ and  ${\widetilde a}_2 = \frac{3}{5}$,
so there is no restriction on unitarity of $\ab{\frac{7}{16}}{b \ge \half}$
and $\ab{\frac{3}{80}}{b \ge \frac{1}{16}}$.

%Highest weight states with $a=\frac{1}{10}$ and $a=\frac{3}{5}$
%can be unitary only if they lie
%on intersection with $d_1=0$: $\ab{\frac{1}{10}}{\frac{1}{10}}$ and
%$\ab{\frac{3}{5}}{0}$. There are no unitary states on $a=\frac{3}{2}$
%line since it doesn't intersect the $d_1=0$ curve in the region $\{ a\ge 0, \, b\ge 0 \}$.
%$\{ a=0, \, b \ge 0 \}$ line lying on $f_{1,1}=0$ stays unitary. 
%States with weights $a=\frac{7}{16}$ and  $a=\frac{3}{80}$
%are from the middle of the Kac table, $\ab{\frac{7}{16}}{b}$ state has \desc s
%on level $\half$ with $\widetilde a = 0$ and  $\widetilde a = \frac{3}{2}$,
%$\ab{\frac{3}{80}}{b}$ has \desc s with  
%$\widetilde a = \frac{1}{10}$ and  $\widetilde a = \frac{3}{5}$,
%so there is no restriction on unitarity from level $\half$ \desc s.

\item[Example 2.] $\quad c=12$

Again the region below $d_1=0$ is nonunitary (Figure~\ref{fig12}).
$\ab{\half}{\half}$ is the only unitary state on $a=\half$ line.
$\ab{\frac{1}{16}}{b \ge\frac{7}{16} }$ has \desc s with  
${\widetilde a} = 0$ and  ${\widetilde a} = \frac{1}{2}$, it remains unitary.
$\ab{0}{b \ge 0}$ is unitary since it lies on $f_{1,1}=0$ line.

\end{description}

\begin{figure}                          
\centering
\includegraphics[width=340pt]{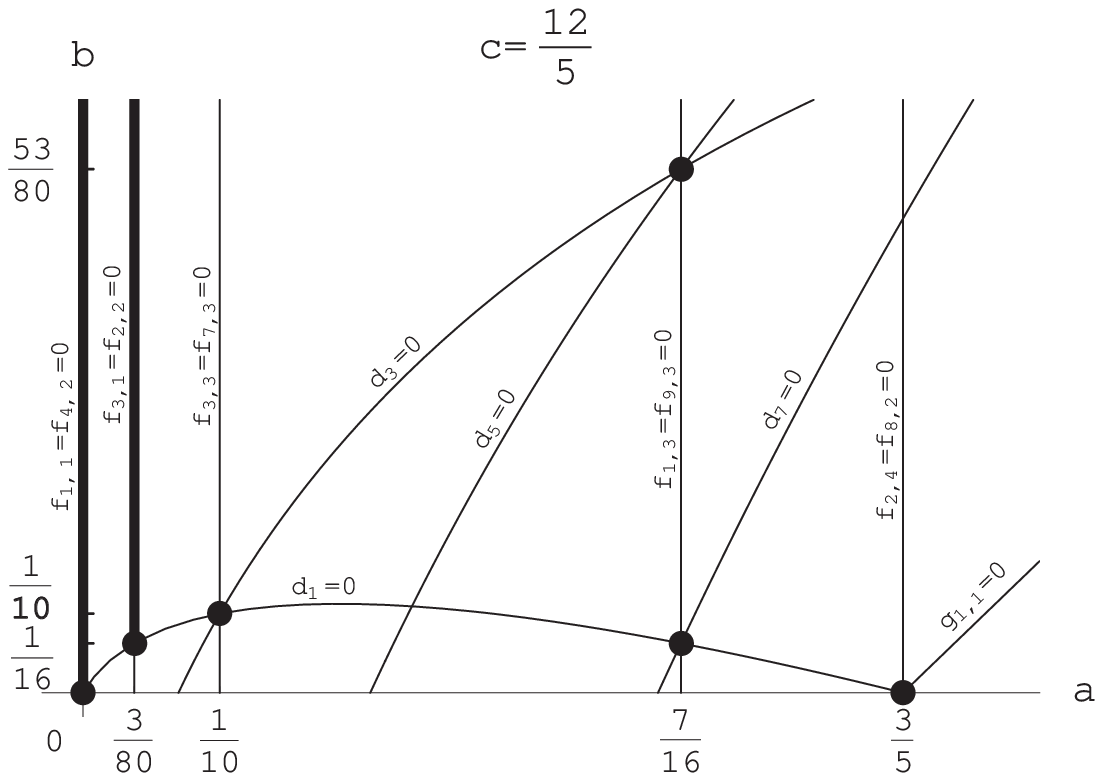}
{\caption{NS sector of $c=12/5$ model.
\small{Thick lines and dots show the continuous and discrete unitary
spectrum respectively, 
thin lines show the
vanishing locus of the Kac determinant.}}   \label{fig125}}
\phantom{nothing}
\includegraphics[width=340pt]{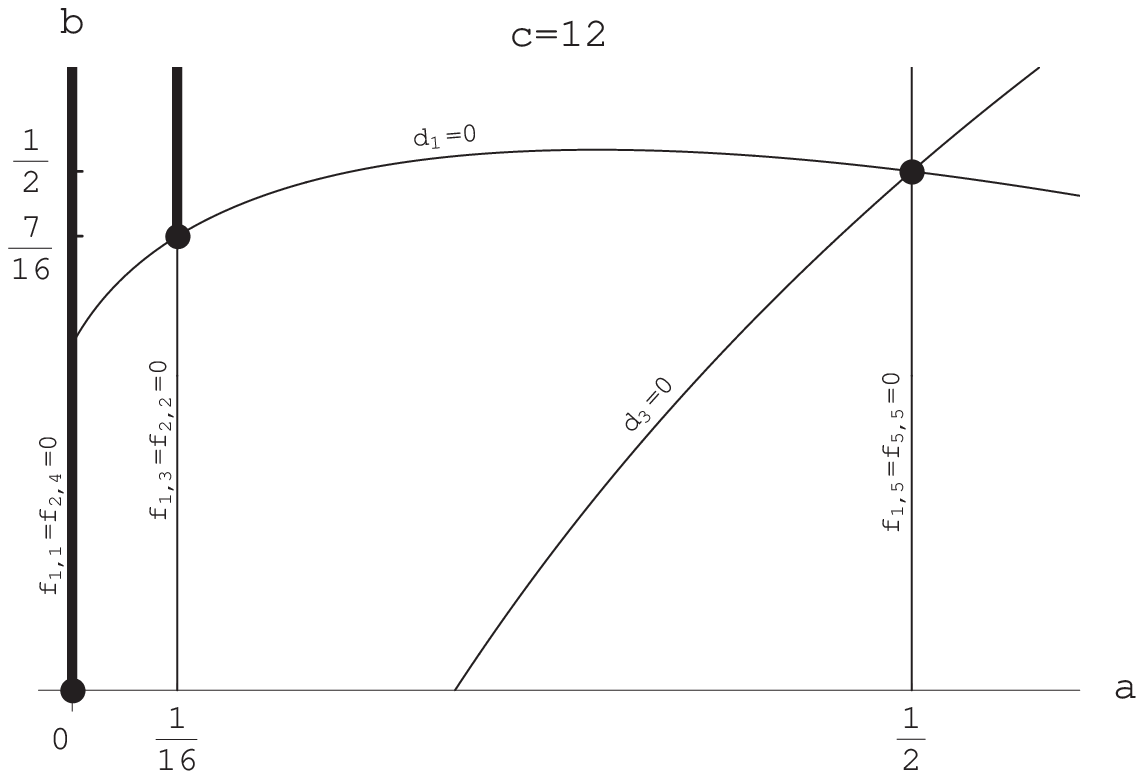}
{\caption{NS sector of $c=12$ model. 
\small{Thick lines and dots show the continuous and discrete unitary
spectrum respectively, 
thin lines show the
vanishing locus of the Kac determinant.}}   \label{fig12}}
%\phantom{nothing}%\\[2ex]
%\centering
%Thick lines and dots show the continuous and discrete unitary
%spectrum respectively, 
%thin lines show the
%vanishing locus of the Kac determinant.
\end{figure}

Now we proceed to the deeper levels. 
Two level 1 \desc s 
\begin{equation}
\barray{rl}
G_{-1/2} \, U_{-1/2} 
- \frac{54\, h +3-2\, c}{\sqrt{15-c}\,\sqrt{21+4 \, c}}\, A_{-1}&
\ab{a}{b} \, ,\\
B_{-1} & \ab{a}{b}
\earray
\end{equation}
are $A$--primaries having the same $A_0$ eigenvalue $a$ as $\ab{a}{b}$.

Level $\frac{3}{2}$ \desc s repeat the situation on level $\half$:
they can be organized to $A$ \hwr s with $A_0$ eigenvalue
being nearest to $a$ neighbor in the row (column) of the Kac table.

Descendants with a new feature appear on level 2. 
Starting from this level one can act twice by the modes of
$G$: $G_{-3/2} \, G_{-1/2} \ab{a_{m,n}}{b}$, producing shift by $2$
in the Kac table by repeated action of $\Phi_{1,2}$ (\ref{shift-fusion}). 
$A_0$ eigenvalues are next to nearest neighbors to $a_{m,n}$
in the same row: $a_{m, n\pm 2}$ (for models in ascending series),
or in the same column: $a_{m\pm 2, n}$ (for models in descending series)
of the correspondent Kac table.
Appearance of these \desc s restricts the unitarity of 
highest weight NS representation $\ab{a}{b}$, if its internal dimension
$a$ belongs to second or last but one column (row) of the Kac table,
since at least one of its new \desc s is from outside of the Kac table.
Again there is one possibility to save the unitarity:
to find such a value of $b$, that the nonunitary \desc\ of $\ab{a}{b}$
becomes a null state. This can happen only
on vanishing locus of det$M^{\mathrm{NS}}_2$, \sw Kac determinant
on level 2.

 As we go to the deeper levels, we get a new shift  in the Kac table on level
$\frac{9}{2}$ (the level of $G_{-5/2} \, G_{-3/2} \, G_{-1/2} \ab{a}{b}$). 
The \desc s of $\ab{a_{m,n}}{b}$ have $A_0$
eigenvalues $a_{m, n\pm 3}$ ($a_{m\pm 3, n}$). One can 
proceed to build new \desc s which are $A$ primaries those
$A_0$ eigenvalue shifted farther in the Kac table. 
Every \hwr\ $\ab{a_{m,n}}{b}$ gets nonunitary \desc\ on large enough level.
To preserve its unitarity the representation have to be 
degenerated on this level in a way, that the nonunitary \desc\ is a null state.

In general, the shift by $k$ in the Kac table appears starting from level
\begin{equation}
\sum_{j=1}^k \left( j-\half \right) = \half \, k^2 \, .
\end{equation}
%We will discuss this formula in the next section,
%but now we return to the examples. 

%\begin{figure}                             
%\centering 
%\includegraphics
%{gr125.eps}
%\caption{NS sector of $c=12/5$ model}                                      \label{fig125}
%%\end{figure}
%%\begin{figure}[h]                               
%\centering 
%\includegraphics
%{gr12.eps}
%\caption{NS sector of $c=12$ model}                                      \label{fig12}
%Thick lines and dots show the continuous and discrete unitary
%spectrum respectively, 
%thin lines show the
%vanishing locus of the Kac determinant.
%\end{figure}

\begin{description}

\item[Example 1.] $\quad c=12/5$

$\ab{\frac{3}{80}}{b}$ would have two nonunitary \desc s on 
level 2: ${\widetilde a}=\frac{63}{80}$ and ${\widetilde a}=\frac{143}{80}$,
but both are null, since $a= \frac{3}{80}$ lies on $f_{3,1}=f_{2,2}=0$, meaning
that there are independent null states on levels $N=\frac{3}{2}$ and $N=2$.

Level 2 nonunitary \desc s of $\ab{\frac{7}{16}}{b}$: 
${\widetilde a}=\frac{3}{16}$ and ${\widetilde a}=\frac{51}{16}$.
One of them is null, since $\ab{\frac{7}{16}}{b}$ lies on 
$f_{1,3}=0 \, \left( N=\frac{3}{2}\right)$, the second is null only 
on intersection with $d_1=0$ or $d_3=0$: $\ab{\frac{7}{16}}{\frac{1}{16}}$
and $\ab{\frac{7}{16}}{\frac{53}{80}}$ respectively.

The full analysis shows that there are no other restrictions 
on unitarity from deeper than  $\half$ levels. As a final result 
we get a picture of unitary representations at $c=12/5$
(Figure~\ref{fig125}).\\

\item[Example 2.] $\quad c=12$

No new unitarity restrictions  from deeper than  $\half$ levels.
Figure~\ref{fig12} represents the unitary spectrum of the model.

\end{description}

The structure of Ramond representations is similar to that of NS representations.
The \desc s are shifted in the row (column) of the Kac table. The
nearest neighbors to the pair of Ramond internal dimensions appear at level 1,
next to  nearest -- at level 3. Shift by $k$ starts from level
\begin{equation}
\sum_{j=1}^k j = \half \, k \,(k+1) .
\end{equation}

We have shown, that every unitary state of \sw algebra is decomposed 
to a sum of representations of internal Virasoro algebra $A$
with internal dimensions belonging to the same row (column)
of the Kac table. Moreover, {\it every} dimension in the row (column)
is present in the decomposition of the unitary state.
In tables~\ref{tab125} and \ref{tab12} we list
unitary states in NS and Ramond sectors of our two example models
and the decomposition of the states to \hwr s of the internal algebra.
\begin{table}[htp]
$$
\setlength\arraycolsep{0.77ex}{
\begin{array}{|c|cc||c|cc|}
\hline
\multicolumn{3}{|c||}{\mathrm{NS}} &  \multicolumn{3}{c|}{\mathrm{Ramond}}\\
\hline
\mathrm{state}&\multicolumn{2}{c||}{\mathrm{decomposition}}&
\mathrm{state}&\multicolumn{2}{c|}{\mathrm{decomposition}}\\
\hline
&&&&&\\[-2.3ex]
\ab{0}{0} & \abi{\frac{7}{16}}{\frac{17}{16}}&\abi{\frac{3}{2}}{\frac{5}{2}}&
\Rab{\frac{7}{16}}{\frac{17}{16}}{\frac{3}{2}}{0} & 
\abi{0}{\frac{5}{2}}& \\[2.1ex]
\ab{\frac{7}{16}}{\frac{1}{16}} & \abi{0}{1}&\abi{\frac{3}{2}}{\frac{1}{2}}&
\Rab{0}{\half}{\frac{7}{16}}{\frac{1}{16}}&\abi{\frac{3}{2}}{1}& \\[2.1ex]
\ab{\frac{7}{16}}{\frac{53}{80}}&\abi{0}{\frac{8}{5}}&\abi{\frac{3}{2}}{\frac{1}{10}}&
\ab{0}{\frac{1}{10}}& \abi{\frac{7}{16}}{\frac{53}{80}}&\abi{\frac{3}{2}}{\frac{8}{5}}\\[0.77ex]
\ab{\frac{1}{10}}{\frac{1}{10}} &\abi{\frac{3}{80}}{\frac{53}{80}}&\abi{\frac{3}{5}}{\frac{8}{5}}&
\Rab{\frac{3}{80}}{\frac{53}{80}}{\frac{3}{5}}{\frac{1}{10}}&\abi{\frac{1}{10}}{\frac{8}{5}}& \\[2.1ex]
\ab{\frac{3}{80}}{\frac{1}{16}} &\abi{\frac{1}{10}}{\frac{1}{2}}&\abi{\frac{3}{5}}{1}&
\ab{\frac{3}{80}}{\frac{1}{16}} &\abi{\frac{1}{10}}{1}&\abi{\frac{3}{5}}{\half} \\[0.77ex]
\ab{\frac{3}{5}}{0} &\abi{\frac{3}{80}}{\frac{17}{16}}&\abi{\frac{1}{10}}{\frac{5}{2}}&
\ab{\frac{1}{10}}{0} &\abi{\frac{3}{80}}{\frac{17}{16}}&\abi{\frac{3}{5}}{\frac{5}{2}}\\[0.77ex]
\hline
&&&&&\\[-2.3ex]
\ab{0}{x} &  \abi{\frac{7}{16}}{\frac{1}{16}\! + \! x}&\abi{\frac{3}{2}}{\frac{1}{2}\! + \! x}&
\Rab{0}{\half\! + \! x}{\frac{7}{16}}{\frac{1}{16}\! + \! x}&\abi{\frac{3}{2}}{x}&  \\[2.1ex]
\ab{\frac{3}{80}}{\frac{1}{16}\! + \! x} &\abi{\frac{1}{10}}{\frac{1}{2}\! + \! x}&\abi{\frac{3}{5}}{x}&
\Rab{\frac{1}{10}}{x}{\frac{3}{80}}{\frac{1}{16}\! + \! x}&
\abi{\frac{3}{5}}{\frac{1}{2}\! + \! x}&\\[2.1ex]
\hline
 \end{array}
}
$$
\caption{Unitary states of $c=12/5$ model and their decomposition.}          \label{tab125}
%\end{table}
%\begin{table}[h]
$$
\setlength\arraycolsep{0.77ex}{
\begin{array}{|c|cc||c|cc|}
\hline
\multicolumn{3}{|c||}{\mathrm{NS}} &  \multicolumn{3}{c|}{\mathrm{Ramond}}\\
\hline
\mathrm{state}&\multicolumn{2}{c||}{\mathrm{decomposition}}&
\mathrm{state}&\multicolumn{2}{c|}{\mathrm{decomposition}}\\
\hline
&&&&&\\[-2.3ex]
\ab{0}{0} & \abi{\frac{1}{16}}{\frac{23}{16}} &\abi{\frac{1}{2}}{\frac{7}{2}}&
\ab{\half}{0} & \abi{\frac{1}{16}}{\frac{23}{16}} &\abi{0}{\frac{7}{2}} \\[0.77ex]
\ab{\frac{1}{16}}{\frac{7}{16}} & \abi{0}{1} & \abi{\frac{1}{2}}{\frac{3}{2}}&
\ab{\frac{1}{16}}{\frac{7}{16}} & \abi{0}{\frac{3}{2}} & \abi{\frac{1}{2}}{1}\\[0.77ex]
\ab{\half}{\half}&\abi{\frac{1}{16}}{\frac{23}{16}} &\abi{0}{3}&
\ab{0}{\half}&\abi{\frac{1}{16}}{\frac{23}{16}} &\abi{\frac{1}{2}}{3}\\[0.77ex]
\hline
&&&&&\\[-2.3ex]
\ab{0}{x} &\abi{\frac{1}{16}}{\frac{7}{16}\! + \! x} &\abi{\frac{1}{2}}{\frac{3}{2}\! + \! x}&
\Rab{\half}{x}{\frac{1}{16}}{\frac{7}{16}\! + \! x} &\abi{0}{\frac{3}{2}\! + \! x}& \\[2.1ex]
\ab{\frac{1}{16}}{\frac{7}{16}\! + \! x}& \abi{0}{1\! + \! x}&\abi{\frac{1}{2}}{\frac{1}{2}\! + \! x}&
\Rab{\frac{1}{16}}{\frac{7}{16}\! + \! x}{0}{\frac{1}{2}\! + \! x}&\abi{\frac{1}{2}}{1\! + \! x}& \\[2.1ex]
\hline
\end{array}
}
$$
\caption{Unitary states of $c=12$ model and their decomposition.}          \label{tab12}
\end{table}
The new notation is introduced: $\abi{a}{b}$ stays for \hws\  of 
$A$ Virasoro algebra, which is eigenstate of $B_0$ operator:
\begin{equation}                                                          \label{abi}
\barray{rrcll}
A_n &\abi{a}{b}&=&0& \qquad n \ge 0  \, ,\\
A_0 &\abi{a}{b}&=&a &\abi{a}{b}  ,\\
B_0 &\abi{a}{b}&=&b &\abi{a}{b}  .
\earray
\end{equation}
When we write $\abi{a}{b}$ we actually mean infinite number 
of states $\abi{a}{b+k}$, $k$ being positive integer or zero.
These states are obtained by action of negative modes of $B$ on $\abi{a}{b}$.
Obviously, $\abi{a}{b+k}$ also satisfies definition (\ref{abi})
with $B_0$ eigenvalue $b+k$.

The lengthly list of all the unitary representations of 
descending and ascending series of models is given in appendix~\ref{apB}.
We would like to point out here some common features of the models.
\begin{itemize}

\item
At fixed $c_p$ there are unitary lines (continuous spectrum)
and separated points (discrete spectrum) in two dimensional space of weights.
We include the points of beginning of unitary lines to discrete spectrum.
Continuous spectrum of each sector consists of $[\frac{p\, (p-2)+1}{4} ]$ lines
at $c=\cl$ and of  $[\frac{(p+1)\, (p-1)+1}{4} ]$ lines at $c=\cu$. Discrete
spectrum consists of $\frac{p \, (p+1)}{2}$ states in each sector at  $c=\cl$
or $c=\cu$.

\item
Every model contains in its discrete spectrum NS state of dimension $h=\half$:
$\ab{a_{1,2}}{\half-a_{1,2}}$ in ascending series and  $\ab{a_{2,1}}{\half-a_{2,1}}$
in descending series. In Section~\ref{section:extensions}
 we extend the \sw algebra to include this state.
In addition every model in descending series contains NS state of dimension $h=1$:
$\ab{a_{3,1}}{\half-a_{3,1}}$. All the models include NS continuous line: $\ab{0}{b \ge 0}$.

\item
%\sloppy
The purely internal states, i.e.
having zero external dimension $b=0$, are important for analysis of fusion rules 
(see the next section). In addition to the vacuum state $\ab{0}{0}$ 
there is always $\ab{a_{3,1}}{0}$  ($\ab{a_{1,3}}{0}$) state in the NS discrete spectrum.
Ramond sector contains ground state $\ab{a_{2,1}}{0}$ ($\ab{a_{1,2}}{0}$)
and two dimensional state
$\Rab{a_{4,2}}{a_{4,1}-a_{4,2}}{a_{4,1}}{0}$  \linebreak[3]
$\left( \Rab{a_{2,4}}{a_{1,4}-a_{2,4}}{a_{1,4}}{0} \right)$.
The existence of these representations was predicted 
in the end of Section~\ref{constr} using relation
(\ref{magic_relation}). 
There were other local or semilocal states $\abi{a_{m,1}}{0}$ 
($\abi{a_{1,n}}{0}$),
but they have higher than $1$ or $\frac{3}{2}$ poles in their \ope\
with $G$, therefore they are se\-con\-dary fields. The correspondent primary fields
are easily found in the discrete spectrum. For example, in the $c=3$
($p=5$) model the NS state $\abi{a_{5,1}=3}{0}$ is level $\half$ 
\desc\ of $\ab{\frac{7}{5}}{\frac{11}{10}}$.
In the $c=24/7$ ($p=6$) model $\abi{a_{5,1}=\frac{22}{7}}{0}$
is level $\half$ \desc\ of $\ab{\frac{85}{56}}{\frac{9}{8}}$
and $\abi{a_{6,1}=5}{0}$ is level 1 \desc\ of 
$\Rab{\frac{4}{3}}{\frac{8}{3}}{\frac{23}{8}}{\frac{9}{8}}$.

\item
There is the same number of states in NS and Ramond sectors, the number
of discrete states is the same and the number of continuous lines is
the same.
%$N=1$ and $N=2$ superconformal models also have this feature. 
We discuss it in the next section.
 
\end{itemize}

%%%%%%%%%%%%%%%%%%%%%%%%%%%%%%%%%%%%%%%

\section{Fusion rules and minimal models}  \label{sec:Fusion-rules-minimal}

\setcounter{equation}{0}

We can not say exactly what are the fusion rules of all fields in the unitary model.
However there are some selection rules induced by the fusion rules of internal
Virasoro minimal model \cite{BPZ}:
%Let's take two states in the ascending series
%model, those internal dimensions belong to the first row of the Kac table.
%The decomposition of these states to representations of the internal algebra
%contains only representations from the same first row.  First row of the Kac table
%is closed under fusion rules, therefore the fusion of these two states
%includes {\it only} states from the first row. It is not hard to extend this logic 
%to the general row and to show, 
\begin{equation}
\Phi_{m_1,n_1} \times \Phi_{m_2,n_2} = 
\sum_{m=|m_1-m_2|+1}^{m_1+m_2-1} \,\,
\sum_{n=|n_1-n_2|+1}^{n_1+n_2-1} \,\,
\Phi_{m,n} \, ,
\end{equation}
where the indices $m$ and $n$ in the sums are raised by steps of 2.
Essentially these fusion rules are two separate selection rules:
one is for the row number,
another is for the column number.
In the case of ascending (descending) series model the column (row)
selection rule has no meaning any more, since any 
unitary \sw representation contains a whole row (column) of the minimal
model Kac table. But another selection rule, row (column) rule is still valid.
In $c=\cl$ model the Kac table row number of the internal dimension
satisfies the $SU(2)$ selection rule:
\begin{equation}                               \label{selectrul}
m \in \{ |m_1-m_2|+1, |m_1-m_2|+3, \ldots ,|m_1+m_2|-1 \}\, .
\end{equation}
(The same rule for column number $n$ in $c=\cu$ model.)
Moreover every row in the set must be represented in the fusion rule.

%In the case of descending series there is the same selection rule (\ref{selectrul})
%on columns of the Kac table.

We have mentioned in the previous section, that every model contains 
a field with zero external dimension. Fusion rules of such a field 
can be written exactly, since it doesn't change the external dimension
of fields, it is acting on.
 Take for example the fusion of $\ab{\frac{3}{5}}{0}$ state of $c=12/5$
model with another state from the unitary spectrum, 
e.g. $\vphantom{\vec{A}}\ab{\frac{1}{10}}{\frac{1}{10}}$:
\begin{equation}
\ab{\frac{3}{5}}{0} \times \ab{\frac{1}{10}}{\frac{1}{10}} =
\abi{\frac{1}{10}}{\frac{1}{10}} + \abi{\frac{3}{2}}{\frac{1}{10}},
\end{equation}
where internal dimensions are obtained according to the fusion rules of 
the Tricritical Ising model. $\abi{\frac{1}{10}}{\frac{1}{10}}$ is of course
simply the $ \ab{\frac{1}{10}}{\frac{1}{10}} $ state, and 
$\abi{\frac{3}{2}}{\frac{1}{10}}$ is found to be the level $\half$
\desc\ of $\ab{\frac{7}{16}}{\frac{53}{80}}$ (see table \ref{tab125}).
Another example from the same model:
\begin{equation}
\Rab{\frac{7}{16}}{\frac{17}{16}}{\frac{3}{2}}{0} \times
\Rab{\frac{1}{10}}{x}{\frac{3}{80}}{\frac{1}{16}\! + \! x}=
\abi{\frac{3}{5}}{x}+\abi{\frac{3}{80}}{\frac{1}{16}\! + \! x}=
\ab{\frac{3}{80}}{\frac{1}{16}\! + \! x} \, .
\end{equation}

Some states have purely internal \desc s, as was shown in the end of
the previous section. Due to this feature we also know all the fusions
of these representations. For example, in the $c=3$ model the fusion
of $\ab{\frac{7}{5}}{\frac{11}{10}}$ with any other field is dictated by its
$\abi{3}{0}$ \desc . In particular, its square is identity:
\begin{equation}
  \ab{\frac{7}{5}}{\frac{11}{10}} \times
  \ab{\frac{7}{5}}{\frac{11}{10}} =
\ab{0}{0} \, .
\end{equation}

Every \sw model contains such a representation in its discrete
spectrum, since every Virasoro minimal model contains 
$\Phi_{p,1}=\Phi_{1,p-1}$ field, square of which is  identity.
From relation (\ref{magic_relation}) we get that such a representation
is of NS type for $p$ odd and of Ramond type for $p$ even
in the ascending series of models and vice versa in the descending series 
of models. This is closely related to the fact, that there is
the same number of states in different sectors.
 
$N=1$ and $N=2$ superconformal minimal models also have equal number of
states in NS and Ramond sectors. In the  $N=2$ case ($c=\frac{3\, k}{k+2}\, ,
\, k=1,2, \ldots $) there is an isomorphism between two sectors preserving the
fusion rule structure. Moreover there exists a continuous transformation,
$U(1)$ flow, connecting the isomorphic states
\cite{Schwimmer:1987mf}. 
In the case of $N=1$
superconformal minimal models 
($ c=\frac{3}{2}-\frac{12}{k\,(k+2)}\, , \, k=3,4,\ldots $)
there is no such a continuous flow, but in every second model ($k=3,5,7,\ldots$)
the isomorphism exists. The isomorphism map is performed by Ramond state,
fusion of which with itself gives the vacuum ($h=\frac{7}{16}$ Ramond state in
the Tricritical Ising model).

In the case of \sw algebra the situation is the same as for $N=1$
superconformal algebra.
\sw algebra admits no continuous transformation between the sectors.
 The isomorphism exists in every second model, exactly then
the state $\abi{a_{p,1}}{0}$ ($\abi{a_{1,p-1}}{0}$) is of Ramond type.
 It serves as the isomorphism map. 
One can find two examples of the 
NS -- Ramond correspondence in Tables \ref{tab125} and 
\ref{tab12}:  the representations in the same row are isomorphic. 

Finally we come to the question, what are the minimal
models of \sw algebra? The states of continuous spectrum
can not be included in the minimal models. Their fusions are not 
defined, since the continuous spectrum representations
are not completely degenerated, they have only
2 independent null vectors. The fusions of discrete
spectrum representations in general also can not be defined,
they depend on the particular realization of the algebra
and may include representations from  continuous spectrum.
For example, in the $c=3/2$ model (\ref{c32}, \ref{c32NS}, \ref{c32R})
one can compactify the free boson on any radius. Then 
the fusion of the boson twist field $s$ with itself depends on
the compactification radius.

However, in every model there is a subset of discrete
spectrum fields with well defined fusion rules, 
and the set is closed under these fusion rules.
It consists of the purely internal representations and the
representations having purely internal \desc s:
$\abi{a_{m,1}}{0}$ ($\abi{a_{1,n}}{0}$).
This is in some sense trivial set of fields, but the only
one, guaranteed to close under fusion rules.
So, the \sw minimal models in addition to the ground states
\begin{equation}
  \text{NS: } a=h=0 \, , \qquad \text{R: } a=h=c/24
\end{equation}
contain at $c=6-\dfrac{18}{p+1}$ (ascending series of minimal models)
$p-2$ fields
\begin{equation}
\begin{aligned}
  a&=\frac{\left( p\,(m+3)-(m+1)-r \, (p+1) \right)^2-4}
{16 \, p\, (p+1)} \, ,
\\
h&=\frac{m^2 \, (p-1)+2 \, m \, (3\, p-1)+p-5}{8\, (p+1)}
+\frac{|r|}{8} \, ,\\
%\end{aligned}
%\qquad
%\begin{gathered}
m&=1,2,\ldots \! , p-2, \qquad 
%\begin{array}{rll}
 \text{NS: }m  \text{ -- odd}, r=0, \quad
\text{R: } m  \text{ -- even},  r=\pm 1 
\end{aligned}
%\end{array}
\end{equation}
and at $c=6+\dfrac{18}{p}$ (descending series of minimal models)
$p-3$ fields
\begin{equation}
\begin{aligned}
  a&=\frac{\left( p\,(n+3)+2\, n+4-r \, p \right)^2-4}
{16 \, p\, (p+1)} \, ,
\\
h&=\frac{n^2 \, (p+2)+2 \, n \, (3\, p+4)+p+6}{8\, p}
+\frac{|r|}{8} \, , \\
%\end{aligned}
%\qquad
%\begin{gathered}
n&=1,2,\ldots \! ,p-3, \qquad
%\begin{array}{rll}
 \text{NS: } n  \text{ -- odd}, r=0, \quad
\text{R: } n  \text{ -- even}, r=\pm 1 .
\end{aligned}
%\end{array}
%\end{gathered}
\end{equation}
One can get the minimal model representations (excluding the vacuum state)
as intersections of $g_{1,1}=0$ and $d_m=0$ vanishing curves of the 
Kac determinant.

There are also nonunitary minimal models of \sw algebra.
They correspond to nonunitary minimal models of internal Virasoro algebra,
central charge of which is 
\begin{equation}
c_A=1- \frac{6 \, (p-q)^2}{p \, q} \, .
\end{equation}
Then the full central charge is 
\begin{equation}
c=-6+18 \, \frac{p-q}{p}  \, .
\end{equation}
The condition of unitarity is $|p-q|\le 1$. 
We do not study the representation theory
of nonunitary minimal models here.

%%%%%%%%%%%%%%%%%%%%%%%%%%%%%%%%%%

\section{Extensions of \sw algebra}           \label{section:extensions}

\setcounter{equation}{0}

In this section we discuss two types of extensions 
of \sw algebra.

\subsection{$\mathcal{SW}(3/2, 3/2, 2)$ algebra}

We have mentioned in the end of Section \ref{unitar} that every
unitary model contains dimension $\half$ field in its discrete spectrum.
It lies on intersection of $f_{1,3}^{\mathrm{NS}}=0$ and $d_1^{\mathrm{NS}}=0$.
For general value of central charge the intersection is 
$\ab{\frac{15-c}{2\,(12+c)}}{\half - \frac{15-c}{2\,(12+c)}}$.
We denote the field corresponding to this state by $\Psi$,
and the correspondent \sw conformal family  -- by $[ \Psi ]$.

We are going to show, that  $[ \Psi ]$ multiplet consists of
4 Virasoro primary fields $\Psi$, $\Phi$, $\Omega$ and $\Sigma$ of dimensions $\half$,
$1$, $\frac{3}{2}$ and $2$ respectively;
and that \sw algebra can be uniquely extended to include this multiplet.
The method, we use, is similar to that of Section \ref{constr} and we omit here
detailed calculations.

Field $\Psi(w)$ obeys \ope s (\ref{opeNS}) with $h=\half$ and $a=\frac{15-c}{2\,(12+c)}$.
$(G_{-1/2} \Psi)$ and $(U_{-1/2} \Psi)$ are not independent due to null vector on
level $\half$:
\begin{equation}
\left( U_{-1/2} - \frac{2\, \sqrt{15-c}}{\sqrt{21+4\,c}} \, G_{-1/2} \right) \Psi =0 \, .
\end{equation}
One new field of dimension 1 has to be introduced:
\begin{equation}
\Phi=G_{-1/2} \Psi \, .
\end{equation}
The field 
\begin{equation}
\Omega=A_{-1}\Psi -\frac{15-c}{12+c} \, \dd \Psi
\end{equation}
is $T$--primary of dimension $\frac{3}{2}$.

The second independent null vector appears on level $\frac{3}{2}$:
\begin{equation}
A_{-1} G_{-1/2} \Psi = A_{-1} U_{-1/2} \Psi =0 \, .
\end{equation}
So the only new field of dimension 2 is $U_{-3/2} \Psi$.
We define
\begin{equation}
\Sigma= \frac{\sqrt{15-c}\, \sqrt{21+4c}}{3\,(12+c)} \, U_{-3/2} \Psi-
\frac{15-c}{12+c}\, L_{-1} G_{-1/2} \Psi\, .
\end{equation}

The fields $A_{-1}^2 \Psi$, $U_{-3/2} G_{-1/2} \Psi$,
$A_{-1} U_{-3/2} \Psi$ enter to the \ope s of \sw generators
with $\Psi$, $\Phi$, $\Omega$, $\Sigma$. But due to \desc s of the two null vectors 
discussed above they can be substituted by composites and derivatives
of the earlier introduced fields:
\begin{equation}
\begin{split}
A_{-1}^2 \Psi &=\frac{18}{12+c} \, A_{-2} \Psi  \, , \\
U_{-3/2} G_{-1/2} \Psi &=- \frac{3\,(12+c)}{\sqrt{15-c}\, \sqrt{21+4c}}
\, L_{-1} A_{-1} \Psi \, , \\
A_{-1} U_{-3/2} \Psi &=-\frac{27}{\sqrt{15-c}\, \sqrt{21+4c}}
A_{-1} G_{-3/2} \Psi \, .
\end{split}
\end{equation}

Using the definitions and the null vectors above and \sw commutation relations
one can write all the \ope s of type $X(z) Y(w)$, where $X$ is one of
\sw generators $\{ T,A,G,U \}$, $Y$ is one of 4 fields, which
constitute the $[ \Psi ]$ multiplet:
$\{ \Psi, \Phi, \Omega, \Sigma \}$.
We do not write these \ope s here because of their length.
One thing to mention: $( \Psi, \Phi )$ and $(\Omega,\Sigma)$ form $h=\half$
and $h=\frac{3}{2}$ multiplets of $N=1$ superconformal algebra
generated by $T$ and $G$.

Now one need to close the extended algebra. The \ope s of type
$Y(z) Y(w)$ are fixed by Jacobi identities  $[ X,Y,Y ]$
and $[Y,Y,Y]$.
The closed algebra, we get, is consistent with fusion
\begin{equation}                             \label{extfus}
[\Psi] \times [\Psi] = [I] \, .
\end{equation}
(No $[\Psi]$ term in the right hand side. $I$ is the identity operator.)

The dimension $\half$ and 1 fields $\Psi$ and $\Phi$ so obtained
are just free fermion and derivative of free boson respectively.
Following the general procedure of \cite{Goddard:1988wv}
the free fermion and the free boson can be decomposed from
the algebra:
\begin{equation}
\begin{split}
T &= {\widetilde T} -\half \NO{\Psi}{\dd \Psi}+ \half \NO{\Phi}{\Phi}
\, ,\\
G &= {\widetilde G} +\NO{\Psi}{\Phi}\, ,\\
%\Omega &= -\frac{3\, \ii \, \sqrt{3+c}}{12+c} \widetilde \Omega\, ,\\
%\Sigma &= -\frac{3\, \ii \, \sqrt{3+c}}{12+c} \widetilde \Sigma\\
A &=\frac{3+c}{12+c} \, {\widetilde A}
 - \NO{\Psi}{ \Omega}
-\frac{15-c}{2\,(12+c)} \NO{\Psi}{\dd \Psi}\, ,\\
U&=\frac{1}{\sqrt{15-c}\, \sqrt{21+4\,c}}
\Big( 2\,(15-c)\,\sqrt{21+4\,c}\, \sqrt{2\,c-3} \, \widetilde U \Big.\\
& \Big. \quad 
+3\,(12+c) ( \NO{\Psi}{\Sigma}-\NO{\Phi}{\Omega}  )+
(15-c) (\NO{\Psi}{\dd \Phi}-2 \NO{\Phi}{\dd \Psi}  ) \Big) \, .
\end{split}
\end{equation}
The operators $\widetilde T$, $\widetilde G$, $\widetilde A$, $\widetilde U$, 
$\Omega$, $ \Sigma$ are commutative with $\Psi$ and $\Phi$,
and form closed operator algebra. The new central charge is 
$\widetilde c= c-3/2$. Conformal dimensions (with respect to $ \widetilde T$)
of generators of the new algebra  are
 $2$, $\frac{3}{2}$, $2$, 
$\frac{5}{2}$, $\frac{3}{2}$, $2$. 
After some redefinition of generators the algebra is found to
coincide with 
$\mathcal{SW}(3/2, 3/2, 2)$  superconformal algebra of Ref.~\cite{Blumenhagen:1992nm}
(for the case of zero coupling $C_{{\mathcal M}{\mathcal M}}^{\mathcal M}$).
At $\widetilde c=10 \half$ the algebra becomes the $G_2$ algebra of Ref.\cite{shv},
the symmetry algebra of manifolds of $G_2$  exceptional holonomy.
The $\mathcal{SW}(3/2, 3/2, 2)$ algebra can be thought as a natural generalization
of the $G_2$ exceptional holonomy algebra to any value of central charge.

We have shown that there is a connection between 
$\mathcal{SW}(3/2,3/2,2)$ and \linebreak[4]  
\sw algebras:
\begin{equation}          \label{g2spin7}
\boxed{
\begin{gathered}
\phantom{\text{\sw}}\\
 \mathcal{SW}(3/2, 3/2, 2) \\
\phantom{h=1/2  \text{ multiplet}}
\end{gathered}
} \otimes
\boxed{
\begin{gathered}
\text{free}\\[-0.77ex] \text{fermion}
\end{gathered}
} \otimes
\boxed{
\begin{gathered}
\text{free}\\[-0.77ex] \text{boson}
\end{gathered}
}
=
\boxed{
\begin{gathered}
\text{\sw} \\
\text{extended by}\\
h=1/2  \text{ multiplet}
\end{gathered}
}
\end{equation}
In the case $\widetilde c=10 \half$, $c=\widetilde c +3/2=12$ 
the tensor product
connects $G_2$ and $Spin(7)$ exceptional algebras discussed in \cite{shv}.

Relation (\ref{g2spin7}) restricts the unitarity of 
$\mathcal{SW}(3/2, 3/2, 2)$ algebra.
Suppose there exists a unitary representation 
of  $\mathcal{SW}(3/2, 3/2, 2)$ algebra at central charge $\widetilde c$.
Multiplying it by some unitary representations of the
free fermion and the free boson theories we get a unitary
representation of extended \sw algebra, which is 
necessarily unitary representation of \sw algebra itself.
This proves that unitary representations of $\mathcal{SW}(3/2, 3/2, 2)$ algebra
can appear only at $\widetilde c = \cl -3/2$ or
$\widetilde c = \cu -3/2$ or $\widetilde c =4 \half$.
The condition is necessary but not sufficient. 
Indeed the theory at $\widetilde c=12/5-3/2=9/10$ is not unitary
by the nonunitarity theorem 
\cite{Friedan:1984xq,Friedan:1986kd}.
The extension of \sw algebra, we got, corresponds to fusion (\ref{extfus}),
which may be inconsistent with unitarity.

We intend to study the representation theory of 
$\mathcal{SW}(3/2, 3/2, 2)$ superconformal
algebra in another paper.

\subsection{$\mathcal{SW}^{2}(1,2)$ algebra}

The second extension we discuss is the extension of 
\sw by $\ab{0}{1}$ representation,
which belongs to continuous spectrum of all  \sw unitary models.
The extended algebra is $N=2$ super--$W_3$ algebra of 
Romans \cite{Romans:1992wi}.
In the conventions of  \cite{Bouwknegt:1993wg}
the algebra is denoted by $\mathcal{SW}^2(1,2)$,
meaning that the algebra consists of $N=2$ superconformal algebra and
its dimension 2 multiplet (4 Virasoro primary fields of dimensions 
$2$, $\frac{5}{2}$,  $\frac{5}{2}$, $3$). Romans has shown that 
$\mathcal{SW}^2(1,2)$
algebra contains \sw as a subalgebra. Here we go opposite way: we
extend \sw to $\mathcal{SW}^2(1,2)$.

$[ J ]$ is \sw conformal family of $\ab{0}{1}$, and $J(w)$ is the field 
corresponding to its highest weight state. It obeys \ope s 
 (\ref{opeNS}) with $h=1$ and $a=0$. The new fields, to be introduced,
correspond to states $G_{-1/2} \ab{0}{1}$, $U_{-3/2} \ab{0}{1}$
and $ U_{-3/2} \, G_{-1/2} \ab{0}{1}$. $A_{-1} \ab{0}{1}$ is null and 
$U_{-1/2} \ab{0}{1}$ is proportional to $G_{-1/2} \ab{0}{1}$.
So the $[ J ]$ multiplet consists of 4 fields of dimensions 
$1$, $\frac{3}{2}$, $\frac{5}{2}$, $3$.
Adding them to the list of dimensions of generators of \sw algebra we get
the list of dimensions of generators of $\mathcal{SW}^2(1,2)$:
$1$, $\frac{3}{2}$, $\frac{3}{2}$, $2$, $2$, $\frac{5}{2}$, $\frac{5}{2}$, $3$.

Unitary models of  $\mathcal{SW}^2(1,2)$ fall to the same values
of central charge $c=\cl$, $c=\cu$, $c=6$ as the unitary models of its
\sw subalgebra.
Romans~\cite{Romans:1992wi} gives coset construction of  $\mathcal{SW}^2(1,2)$
unitary minimal models by
$\frac{SU(3)}{SU(2) \times U(1)}$ Kazama--Suzuki cosets 
\cite{Kazama:1989qp,Kazama:1989uz}.
The ascending series models are represented by coset
\begin{equation}                            \label{coset1}
\frac{SU(3)_k \times SO(4)_1 }
{SU(2)_{k+1} \times U(1)} \, ,
\end{equation}
and the descending  series models are represented by
\begin{equation}                           \label{coset2}
\frac{SU(2,1)_{-k-5} \times SO(4)_1 }
{SU(2)_{-k-4} \times U(1)} \, .
\end{equation}
%The models are necessarily unitary with respect to \sw
%subalgebra. However they are not minimal models of \sw algebra.
% Every $\mathcal{SW}^2(1,2)$ \hwr\ is decomposed to infinite number
%of \sw \hwr s. The only exception is the $c=3/2$ case, which is minimal with
%respect to both algebras.
%
$SO(4)_1$ is decomposed to $SU(2)_1 \times SU(2)_1$.
Then $SU(2)$ in the denominator is given by diagonal 
embedding to the direct sum of $SU(2)$ subalgebra
of $SU(3)$, generated by 
$\{ H^{1}, E^{\alpha_1+\alpha_2}, E^{-\alpha_1-\alpha_2} \}$,
and the first $SU(2)$ in the $SO(4)$.
(The embedding obviously gives the internal Virasoro algebra.)
The generator of $U(1)$  in the denominator is
$H^2 + \sqrt{3} \, J$, 
where $H^2$ is the second Cartan generator of $SU(3)$,
commuting with $E^{\alpha_1+\alpha_2}$, and $J$ is the 
Cartan generator of the second $SU(2)$ in the $SO(4)$.

It is noted in~\cite{Romans:1992wi} that there exists a $\mathbb{Z}_2$
involution of $\mathcal{SW}^2(1,2)$ algebra (see (3.27) 
of~\cite{Romans:1992wi}). The generators of $\mathcal{SW}^2(1,2)$,
invariant under the involution, constitute the \sw subalgebra 
of $\mathcal{SW}^2(1,2)$.
The field content of the $\mathcal{SW}^2(1,2)$ minimal models 
(\ref{coset1}, \ref{coset2})
can be projected by the involution.
One takes a $\mathbb{Z}_2$ invariant combination 
of  $\mathcal{SW}^2(1,2)$ \hws s and decomposes
it to a sum of the \hwr s of \sw algebra. The projected models
represent a nontrivial realization of  \sw symmetry algebra.
Every such a model contains the discrete spectrum 
of \sw algebra and infinite number of \sw representations 
from the continuous spectrum.

%%%%%%%%%%%%%%%%%%%%%%%%%%%%%%%%%%
%\appendix{{\Large {\bf{Appendix}}}}
\appendix
%\section*{Appendix}
%%%%%%%%%%%%%%%%%%%%%%%%%%%%%%%%%%
 
\section{Operator product expansions of \sw \\ algebra}                   \label{apA}

\setcounter{equation}{0}

\begin{align}    
W&=\frac{3\,\left( 12 + c \right) \, A -
    \left( 15 - c \right) \,T}{{\sqrt{15 - 
        c}}\,{\sqrt{21 + 4\,c}}} \, , \nonumber\\
%
%\begin{align}                                
\label{TAA}
T(z)\, A(w)&=A(z)\, A(w)=
\frac{\frac{c \,(15-c)}{6 \,(12+c)}}{(z-w)^4}+\frac{2\,
  A(w)}{(z-w)^2}+\frac{\dd A(w)}{z-w} \, , \\
 \label{AG}
A(z)\, G(w)&= \frac{15-c}{3\, (12 + c)}\left(
\frac{ \frac{3}{2}\, G(w)}{(z-w)^2}+
 \frac{\dd G(w)-\frac{\sqrt{21+4\, c}}{\sqrt{15-c}}\,U(w)}{z-w}
\right) , \\
 \label{AU}
%\shoveleft{
A(z)\, U(w)&={ \frac{1}{(z-w)^3}\left(
-\frac{\sqrt{15  -  c}\,\sqrt{21  +  4\,c}}{12  +  c}\, G(w) \right)+}  
\nonumber\\
& \hspace{-5ex} \frac{1} {(z-w)^2}\left(
\frac{87  +  5\,c}{6\,(12  +  c)}\,U(w)  -  
\frac{\sqrt{15  -  c}\,\sqrt{21  +  4\,c}}{3\,(12  +  c)}\, \dd G(w)
\right)+ \nonumber\\
&\hspace{-5ex} {\frac{1}{z-w} \left( 
\frac{(15  -  c)^{3/2}\, \dd^2 G(w)}{3\,(12  +  c)\, \sqrt{21  +  4\,c}}  
 -  \frac{(15  -  c)\,\dd U(w) }{3\, (12   +   c)}
-\frac{54\, \NO{A}{G}(w)}{\sqrt{15  -  c}\,\sqrt{21  +  4\,c}} 
\right) } , \\
%\raisetag{12ex}
 \label{GG}
G(z)\, G(w)&=\frac{2\,c/3}{(z-w)^3}+\frac{2\, T(w)}{z-w}
\, ,\\
  \label{GU}
G(z)\, U(w)&=\frac{\sqrt{15-c}}{\sqrt{21+4\,c}} \left(
\frac{\frac{12\,(12+c)}{15-c}\, A(w)-4\, T(w)}{(z-w)^2}+
\frac{\frac{3\,(12+c)}{15-c}\, A(w)- T(w)}{z-w}
\right)  , \\
  \label{UU}
 U(z)\, U(w)&=-\frac{2\,c}{(z-w)^5}+\nonumber \\
&\hspace{-5ex} \frac{1}{(z-w)^3}\left(
-\frac{2\,\left( 93 + 10\,c \right) }{21 + 4\,c}\, T(w)-
\frac{12\,\left( 12 + c \right) 
    \left( 6 + 5\,c \right) }{\left( 15 - 
      c \right) \left( 21 + 4\,c \right) }\, A(w) \right)+\nonumber \\
&\hspace{-5ex} \frac{1}{(z-w)^2}\left(
-\frac{93 + 10\,c }{21 + 4\,c}\, \dd T(w)-
\frac{6\,\left( 12 + c \right) 
    \left( 6 + 5\,c \right) }{\left( 15 - 
      c \right) \left( 21 + 4\,c \right) }\, A(w) \right)+\nonumber \\
&\hspace{-5ex} \frac{1}{z-w}\left(
-\frac{3\,\left( 12 + c \right) }{21 + 4\,c}\, \dd^2 T(w)+
\frac{9\,\left( 6 - c \right) \,
    \left( 12 + c \right) }{\left( 15 - c
      \right) \,\left( 21 + 4\,c \right) }\, \dd^2 A(w) -
  \right. \nonumber \\
&\hspace{-5ex} \left. \frac{324\,\left( 12 + c \right) \, \NO{T}{A}(w)}
  {\left( 15 - c \right) \,
    \left( 21 + 4\,c \right) }+
\frac{27\, \NO{G}{\dd G}(w)}{21+4\, c}+
\frac{54\, \NO{G}{U}(w)}{\sqrt{15-c}\,\sqrt{21+4\,c}}
\right) .
%\raisetag{12ex}
\end{align}

%%%%%%%%%%%%%%%%%%%%%%%%%%%%%%%%%%%%%%%%%%%%%%%%%%%%%%%%%%%%%%%%%%%

\section{List of unitary representations of \sw\\ algebra}                            \label{apB}

\setcounter{equation}{0}

$r= \pm 1$ distinguish between $\ket{\uparrow } $ and $\ket{\downarrow } $ states
of two dimensional Ramond representations. 
$x$ stands for real
positive number in the continuous spectrum expressions.
%\begin{center}
\subsection{The unitary spectrum at $c=6- {\dis \frac{18}{p+1}}\,$}
%\end{center}

\noindent {\bf NS sector}

  Discrete spectrum
%\begin{equation}
$$
\barray{|c|c|c|}
\hline 
a & h & m, n \in \mathbb{Z} \\
\hline 
\frac{\left( p \, m - (p + 1)\, n \right)^2 -1}{4\, p\, (p + 1)} &
\frac{p \, (n-m)^2 + (m - 3 \, n+1 )(m - n-1) - m}{2\, (1+p)}&
\begin{array}{c}
1 \le m \le p-1  \\ m \le n \le  \frac{p-1+m}{2} 
\end{array}
\\[1.8ex]
\hline
\frac{\left( p \, m - (p + 1)\, n \right)^2 -1}{4\, p\, (p + 1)} &
\frac{p \, (n-m)^2 + (-m + 3 \, n+1 )(-m + n-1) + m}{2\, (1+p)}&
\barray{c}
 1 \le m \le p-2 \\ m+1 \le n \le  \frac{p+m}{2} 
\earray
\\[1.8ex]
\hline
\earray
$$
%\end{equation}

Continuous spectrum
%\begin{equation}
$$
\barray{|c|c|c|}
\hline 
a & h & m, n \in \mathbb{Z} \\
\hline 
\frac{\left( p \, m - (p + 1)\, n \right)^2 -1}{4\, p\, (p + 1)} &
\frac{p \, (m - n)^2 + (m - 3 \, n+1 )(m - n-1) - m}{2\, (1+p)}+x &
\barray{c}
1 \le m \le p-2  \\ m \le n \le \frac{p-2+m}{2} 
\earray
\\[1.8ex]
\hline
\earray
$$
%\end{equation}

\noindent {\bf Ramond sector}

Discrete spectrum
$$%\begin{equation}
\barray{|c|c|c|}
\hline 
a & h & m, n \in \mathbb{Z} \\
\hline 
\frac{m^2-1}{4 \,p \,(p+1)} &
\frac{p - 2}{4 \, (p+1)}&
1 \le m \le p-1\\
\hline
\frac{\left( p \, m - (p + 1)\, \left( n+ \frac{1-r}{2}\right)
  \right)^2   -  1}{4\, p\, (p + 1)} &
\frac{p \, (n-m)^2 + (n-m)(1 - m + 3 n)+ p \, (n - m) + p/2 - 1}{2 (1+ p)}&
\barray{c}
1 \le   m   \le   p-3 \\ m + 1   \le   n   \le   
\frac{p   -  1  +m}{2} 
\earray
\\[1.8ex]
\hline
\frac{\left( p \, m - (p + 1)\, \left( n- \frac{1+r}{2}\right)
  \right)^2   -  1}{4\, p\, (p + 1)} &
\frac{p \, (n-m)^2 - (n-m)(1 +m - 3 n)- p \, (n - m) + p/2 - 1}{2 (1+ p)} &
\barray{c}
1  \le   m   \le   p-2 \\ m + 1   \le   n   \le   \frac{p+m}{2}
 \earray
\\[1.8ex]
\hline
\earray
$$%\end{equation}

Continuous spectrum
$$%\begin{equation}
\barray{|c|c|c|}
\hline 
a & h & m, n \in \mathbb{Z} \\
\hline 
\frac{\left( p \, m - (p + 1) \left( n- \frac{1+r}{2} \right)  
  \right)^2   -   1}{4\, p\, (p + 1)} &
\barray{c}
\frac{p \, (n-m)^2 - (n-m)(1 +m - 3 n)- p \, (n - m) + p/2 - 1}{2 (1+p)}
+x 
\earray
&
\barray{c}
1  \le   m   \le   p-3 \\ m  +  1   \le   n   \le   \frac{p-1+m}{2} 
\\
\mathrm{and} \\2   \le   m=n   \le   \frac{p+1}{2} 
 \earray
\\[4.5ex]
\hline
\earray
$$%\end{equation}

%\vspace{2ex} 
%\begin{center}
\subsection{The unitary spectrum at $c= 6+{\dis \frac{18}{p}}\, $}
%\end{center}

\noindent {\bf NS sector}

Discrete spectrum
$$%\begin{equation}
\barray{|c|c|c|}
\hline 
a & h & m, n \in \mathbb{Z} \\
\hline 
0 & 0 & \\
\hline
\frac{\left( p\, (n+1)+2\, n+1 \right)^2 -1}{4\, p\, (p + 1)} &
\frac{p\, (n+1)^2 +2\,n \, (n+1)-2\,p}{2\, p}&
1 \le n \le \frac{p}{2}-1\\
\hline
\frac{\left( p \, m - (p + 1)\, n \right)^2 -1}{4\, p\, (p + 1)} &
\frac{p\, (n-m)^2 +(n-m)(2\, m-1)+m+1}{2\, p} &
\barray{c}
1 \le n \le p-2 \\ n+1 \le m \le  \frac{p+n}{2} 
\earray
\\[1.8ex]
\hline
\frac{\left( p \, m - (p + 1)\, n \right)^2 -1}{4\, p\, (p + 1)} &
\frac{p\, (n-m)^2 +(-n+m)(-2\, m-1)-m+1}{2\, p} &
\barray{c}
1 \le n \le p-1 \\  \frac{n}{2} \le m \le  n-1
\earray
\\[1.8ex]
\hline
\earray
$$%\end{equation}

Continuous spectrum
$$%\begin{equation}
\barray{|c|c|c|}
\hline 
a & h & m, n \in \mathbb{Z} \\
\hline 
0  & x & \\
\hline
\frac{\left( p\, (n-1)+2\, n-1 \right)^2 -1}{4\, p\, (p + 1)} &
\frac{(n-1)(p\, (n-1)+2\, n)}{2\, p}+x &
2 \le n \le \frac{p}{2}\\
\hline
\frac{\left( p \, m - (p + 1)\, n \right)^2 -1}{4\, p\, (p + 1)} &
\frac{p\, (n-m)^2 +(n-m)(2\, m-1)+m+1}{2\, p}+x &
\barray{c}
1 \le n \le p-2 \\  n+1 \le m \le \frac{p+n}{2} 
\earray
\\[1.8ex]
\hline
\earray
$$%\end{equation}

\noindent {\bf Ramond sector}

Discrete spectrum
$$%\begin{equation}
\barray{|c|c|c|}
\hline 
a & h & m, n \in \mathbb{Z} \\
\hline 
&&\\[-2.5ex]
\frac{p+3}{4\,p} & \frac{p+3}{4\,p} & \\[0.6ex]
\hline
&&\\[-2.5ex]
\frac{n^2-1}{4\, p\, (p + 1)} &
\frac{p+3}{4\,p}&
1 \le n \le p-1\\[0.6ex]
\hline
\frac{\left(   p (m + \frac{1-r}{2})- (p + 1) n    \right)^2   -  1}{4\, p\, (p + 1)} &
\frac{p \, (m-n)^2 -2\, m\, (m-n)+ p \, (m-n) + p/2 +3/2}{2\, p} &
\barray{c}
1   \le   n   \le   p-3  \\ n  +  1   \le   m   \le   \frac{p-1+m}{2} 
\earray
\\[1.8ex]
\hline
\frac{\left(   p (m + \frac{1-r}{2})- (p + 1) n    \right)^2   -  1}{4\, p\, (p + 1)} &
\frac{p  (m-n)^2 +2 (m-1)(1 - m + n)+ p  (m-n) -3
  (p-1)/2}{2\,p} &
\barray{c}
1   \le   n   \le   p-3  \\ n  +  2   \le   m   \le   \frac{p+1+m}{2} 
\earray
\\[1.8ex]
\hline
\earray
$$%\end{equation}

Continuous spectrum
$$%\begin{equation}
\barray{|c|c|c|}
\hline 
a & h & m, n \in \mathbb{Z} \\
\hline 
\frac{\left( p \, (n + \frac{1-r}{2})+ 2 \, n  \right)^2 -1}{4\, p\,
  (p + 1)}&
\frac{(2\,n-1)(2\,n+1-p)+2 \, p\, n^2 }{4\,p}+x&
1 \le n \le \frac{p-1}{2} \\
\hline
\frac{\left( p \, (m + \frac{1-r}{2})- (p + 1)\, n  \right)^2  -  1}{4\, p\, (p + 1)} &
\frac{p \, (m-n)^2 -2\, m\, (m-n)+ p \, (m-n) + p/2 +3/2}{2\, p}+x &
\barray{c}
1   \le   n   \le   p-3  \\ n  +  1   \le   m  \le   \frac{p-1+m}{2} \\
\mathrm{and} \\
1   \le   n=m   \le   \frac{p}{2}
\earray
\\[4.5ex]
\hline
\earray
$$%\end{equation}

%%%%%%%%%%%%%%%%%%%%%%%%%%%%%%%%%%%%%

\section{Normal ordered product conventions}                   \label{apC}

\setcounter{equation}{0}

The normal ordered product is defined as the zero order term 
in \ope :
\begin{equation}
P(z) \, Q(w)=\mathrm{singular\ terms}+\NO{P}{Q}(w)+O(z-w) \, .
\end{equation}
The mode expansion of $\NO{P}{Q}$ can be easily calculated:
\begin{equation}                       \label{NOexpansion}
  \NO{P}{Q}_n =\sum_{m \le -\Delta_P} P_m Q_{n-m}+
(-1)^{PQ} \! \! \! \! \! \sum_{m \ge -\Delta_P +1} Q_{n-m} P_m \, ,
\end{equation}
where $\Delta_P$ is the dimension of field $P(z)$ and $(-1)^{PQ}$ is
equal to $-1$ only if both $P$ and $Q$ are fermionic operators.
The expansion is valid for   $P$ bosonic  ($m \in \mathbb{Z}$) or   $P$ fermionic
of NS type ($m \in \mathbb{Z}+\half$).
If $P$ is fermionic of Ramond type, then $m \in \mathbb{Z}$ but
$\Delta_P$
is half--integer and uncertainty appears in the limits of the sums in
(\ref{NOexpansion}).

If $P$ is fermionic of Ramond type and $Q$ is bosonic, 
one can express $\NO{P}{Q}$  through $\NO{Q}{P}$ and then
expand by use of (\ref{NOexpansion}).
The problem appears then both $P$ and $Q$ are fermionic
operators of Ramond type. The limits in the sums in  the $\NO{P}{Q}_n$
expansion should be chosen consistently.  We rather fix the limits,
but add some terms to the expansion:
\begin{equation}   
   \NO{P}{Q}_n^{\text{R}} =\! \! \! \sum_{m \le -\Delta_P+1/2} \! \! \! P_m Q_{n-m}-
 \! \! \! \sum_{m \ge -\Delta_P +3/2} \! \! \! Q_{n-m} P_m +
\sum_i \gamma^{(i)}(c,n) \, S_n^{(i)} \, ,
\end{equation}
where $n,m \in \mathbb{Z}$, $S^{(i)}$ are operators entering to the singular part
of \ope\ $P(z) \, Q(w)$, and $\gamma^{(i)}$ are some functions of the
central charge and index $n$.

In the case of \sw algebra we are interested in $\NO{G}{U}$
and $\NO{G}{\dd G}$ expansions:
\begin{align}
\label{NOGU}
 \NO{G}{U}_n^{\text{R}}&=
\sum_{m \le-1}G_m U_{n-m} - \sum_{m \ge 0} U_{n-m} G_m
+\gamma^{(1)} W_n \, ,\\
\label{NOGG}
 \NO{G}{\dd G}_n^{\text{R}}&=
\sum_{m \le-1}G_m (\dd G)_{n-m} - \sum_{m \ge 0} (\dd G)_{n-m} G_m+
\gamma^{(2)}  \delta_{n,0} +\gamma^{(3)} L_n \, .
\end{align}
To fix the coefficients $\gamma$ one can discuss commutator 
$\com{P_n}{\NO{T}{Q}_0}$.
The commutator can be calculated  in two ways.
The first is to calculate the \ope\ $P(z) \, \NO{T}{Q}(w)$, which will include 
$\NO{P}{Q}(w)$. After that the commutator is obtained from the \ope\
by usual procedure. The second way is to expand first $\NO{T}{Q}_0$ by use
of (\ref{NOexpansion}), and then to calculate the commutator.
By equating the results of two computations one fixes the expansion of
$\NO{P}{Q}_n^{\text{R}}$.
Sometimes it is easier to use instead of $T$ another bosonic operator.
For example, we get the $\NO{G}{U}_n^{\text{R}}$ expansion
from $\com{G_n}{\NO{W}{G}_0}$ and the  $\NO{G}{\dd G}_n^{\text{R}}$ 
expansion from $\com{G_n}{{\NO{T}{G}}_0}$.
The relevant \ope s are:
\begin{align}
G(z)\, \NO{W}{G}(w) &= 
\frac{\left( \frac{2}{3}\, c -4 \right) W(w) }{(z-w)^3}-
\frac{3 \, \dd W(w) }{(z-w)^2}  + \nonumber \\
&\hspace{20ex} +\frac{2 \, \NO{T}{W}(w) - \dd^2 W(w) - \NO{G}{U}(w) }{z-w} \, ,\\
G(z)\, \NO{T}{G}(w)&=
\frac{3\, c}{(z-w)^5}+
\frac{\left( \frac{2}{3}\, c +5 \right) T(w)}{(z-w)^3}
+\frac{\frac{3}{2} \, \dd T(w)}{(z-w)^2}  + \nonumber 
\\&\hspace{20ex}+
\frac{2\, \NO{T}{T}(w)+\half \, \dd^2 T(w)-\half \, \NO{G}{\dd
    G}(w)}{z-w} \, .
\end{align}
And the $\gamma$ coefficients in (\ref{NOGU}) and (\ref{NOGG}):
\begin{equation}
  %\begin{split}
\gamma^{(1)}  = \frac{3+n}{2}\, ,  \qquad
\gamma^{(2)}  = \frac{5 \, c}{64} \, , \qquad
\gamma^{(3)}  = \frac{9}{4}+n \, .
%\end{split}
\end{equation}

%%%%%%%%%%%%%%%%%%%%%%%%%%%%%%%%%%%%%
%%%%%%%%%%%%%%%%%%%%%%%%%%%%%%%%%%%%%


\begin{thebibliography}{99}

\bibitem{Zamolodchikov:1985wn}
A.~B.~Zamolodchikov,
``Infinite Additional Symmetries In Two-Dimensional Conformal Quantum Field Theory,''
Theor.\ Math.\ Phys.\ {\bf 65} (1985) 1205.
%%CITATION = TMPHA,65,1205;%%

%\cite{Bouwknegt:1993wg}
\bibitem{Bouwknegt:1993wg}
P.~Bouwknegt and K.~Schoutens,
``W symmetry in conformal field theory,''
Phys.\ Rept.\ {\bf 223} (1993) 183
[hep-th/9210010].
%%CITATION = HEP-TH 9210010;%%

%\cite{Fateev:1987vh}
\bibitem{Fateev:1987vh}
V.~A.~Fateev and A.~B.~Zamolodchikov,
``Conformal Quantum Field Theory Models In Two Dimensions Having Z(3) Symmetry,''
Nucl.\ Phys.\ {\bf B280} (1987) 644.
%%CITATION = NUPHA,B280,644;%%

%\cite{Fateev:1988zh}
\bibitem{Fateev:1988zh}
V.~A.~Fateev and S.~L.~Lukyakhov,
``The Models Of Two-Dimensional Conformal Quantum Field Theory With Z(N) Symmetry,''
Int.\ J.\ Mod.\ Phys.\ {\bf A3} (1988) 507.
%%CITATION = IMPAE,A3,507;%%

%\cite{Lukyanov:1990tf}
\bibitem{Lukyanov:1990tf}
S.~L.~Lukyanov and V.~A.~Fateev,
``Physics Reviews: Additional Symmetries And Exactly Soluble Models 
In Two-Dimensional Conformal Field Theory,''
{\it  Chur, Switzerland: Harwood (1990) 117 p. (Soviet Scientific Reviews A, Physics: 15.2)}.

%\cite{BPZ}
\bibitem{BPZ}
A.~A.~Belavin, A.~M.~Polyakov and A.~B.~Zamolodchikov,
``Infinite conformal symmetry in two-dimensional quantum field theory,''
Nucl.\ Phys.\ {\bf B241} (1984) 333.
%%CITATION = NUPHA,B241,333;%%

\bibitem{fofs}
J.~M.~Figueroa-O'Farrill and S.~Schrans,
``The Conformal bootstrap and super W algebras,''
Int.\ J.\ Mod.\ Phys.\ A {\bf 7} (1992) 591.
%%CITATION = IMPAE,A7,591;%%

%\cite{Figueroa-O'Farrill:1991zg}
%%\bibitem{Figueroa-O'Farrill:1991zg}
J.~M.~Figueroa-O'Farrill and S.~Schrans,
``Extended superconformal algebras,''
Phys.\ Lett.\ {\bf B257} (1991) 69.
%%CITATION = PHLTA,B257,69;%%

\bibitem{shv}
S.~L.~Shatashvili and C.~Vafa,
``Superstrings and manifold of exceptional holonomy,''
hep-th/9407025.
%%CITATION = HEP-TH 9407025;%%

\bibitem{shv2}
S.~L.~Shatashvili and C.~Vafa,
``Exceptional magic,''
Nucl.\ Phys.\ Proc.\ Suppl.\ {\bf 41} (1995) 345.
%%CITATION = NUPHZ,41,345;%%

%\cite{Blumenhagen:1992nm}
\bibitem{Blumenhagen:1992nm}
R.~Blumenhagen,
``Covariant construction of N=1 super W algebras,''
Nucl.\ Phys.\ {\bf B381} (1992) 641.
%%CITATION = NUPHA,B381,641;%%

%\cite{Romans:1992wi}
\bibitem{Romans:1992wi}
L.~J.~Romans,
``The N=2 super W(3) algebra,''
Nucl.\ Phys.\ {\bf B369} (1992) 403.
%%CITATION = NUPHA,B369,403;%%

%\cite{Thielemans:1991uw}
\bibitem{Thielemans:1991uw}
K.~Thielemans,
``A Mathematica package for computing operator product expansions,''
Int.\ J.\ Mod.\ Phys.\ {\bf C2} (1991) 787.
%%CITATION = IMPAE,C2,787;%%

\bibitem{Figueroa-O'Farrill:1997hm}
J.~M.~Figueroa-O'Farrill,
``A note on the extended superconformal algebras associated with  manifolds of exceptional holonomy,''
Phys.\ Lett.\ {\bf B392} (1997) 77
[hep-th/9609113].
%%CITATION = HEP-TH 9609113;%%

%\cite{Friedan:1984xq}
\bibitem{Friedan:1984xq}
D.~Friedan, Z.~Qiu and S.~Shenker,
``Conformal Invariance, Unitarity And Two-Dimensional Critical Exponents,''
in {\it C83-11-10.1}
Phys.\ Rev.\ Lett.\ {\bf 52} (1984) 1575.
%%CITATION = PRLTA,52,1575;%%

%\cite{Friedan:1986kd}
\bibitem{Friedan:1986kd}
D.~Friedan, S.~Shenker and Z.~Qiu,
``Details Of The Nonunitarity Proof For Highest Weight Representations Of The Virasoro Algebra,''
Commun.\ Math.\ Phys.\ {\bf 107} (1986) 535.
%%CITATION = CMPHA,107,535;%%

%\cite{Fradkin:1992bz}
\bibitem{Fradkin:1992bz}
E.~S.~Fradkin and V.~Y.~Linetsky,
``Results of the classification of superconformal algebras in two-dimensions,''
Phys.\ Lett.\ {\bf B282} (1992) 352
[hep-th/9203045].
%%CITATION = HEP-TH 9203045;%%

%\cite{Fradkin:1992km}
\bibitem{Fradkin:1992km}
E.~S.~Fradkin and V.~Y.~Linetsky,
``Classification of superconformal and quasisuperconformal algebras in two-dimensions,''
Phys.\ Lett.\ {\bf B291} (1992) 71
[hep-th/9207035].
%%CITATION = PHLTA,B291,71;%%

%\cite{Ginsparg:1988ui}
\bibitem{Ginsparg:1988ui}
P.~Ginsparg,
``Applied Conformal Field Theory,''
HUTP-88-A054
{\it Lectures given at Les Houches Summer School in Theoretical
  Physics, Les Houches, France, Jun 28 - Aug 5, 1988}.

%\cite{Komata:1991cb}
\bibitem{Komata:1991cb}
S.~Komata, K.~Mohri and H.~Nohara,
``Classical and quantum extended superconformal algebra,''
Nucl.\ Phys.\ {\bf B359} (1991) 168.
%%CITATION = NUPHA,B359,168;%%

%\cite{Mallwitz:1995hh}
\bibitem{Mallwitz:1995hh}
S.~Mallwitz,
``On SW minimal models and N=1 supersymmetric quantum Toda field theories,''
Int.\ J.\ Mod.\ Phys.\ {\bf A10}, 977 (1995)
[hep-th/9405025].
%%CITATION = HEP-TH 9405025;%%

%\cite{Feigin:1982st}
\bibitem{Feigin:1982st}
B.~L.~Feigin and D.~B.~Fuks,
``Invariant Skew Symmetric Differential Operators 
On The Line And Verma Modules Over The Virasoro Algebra,''
Funct.\ Anal.\ Appl.\ {\bf 16} (1982) 114.
%%CITATION = FAAPB,16,114;%%

%\cite{Bershadsky:1985dq}
\bibitem{Bershadsky:1985dq}
M.~A.~Bershadsky, V.~G.~Knizhnik and M.~G.~Teitelman,
``Superconformal Symmetry In Two-Dimensions,''
Phys.\ Lett.\ {\bf B151} (1985) 31.
%%CITATION = PHLTA,B151,31;%%


%\cite{Mussardo:1987eq}
\bibitem{Mussardo:1987eq}
G.~Mussardo, G.~Sotkov and M.~Stanishkov,
``Ramond Sector Of The Supersymmetric Minimal Models,''
Phys.\ Lett.\ {\bf B195} (1987) 397.
%%CITATION = PHLTA,B195,397;%%

%\cite{Mussardo:1989av}
\bibitem{Mussardo:1989av}
G.~Mussardo, G.~Sotkov and M.~Stanishkov,
``N=2 Superconformal Minimal Models,''
Int.\ J.\ Mod.\ Phys.\ {\bf A4} (1989) 1135.
%%CITATION = IMPAE,A4,1135;%%

%\cite{Dotsenko:1984nm}
\bibitem{Dotsenko:1984nm}
V.~S.~Dotsenko and V.~A.~Fateev,
``Conformal algebra and multipoint correlation functions in  2D statistical models,''
Nucl.\ Phys.\ {\bf B240} (1984) 312.
%%CITATION = NUPHA,B240,312;%%

%\cite{Schwimmer:1987mf}
\bibitem{Schwimmer:1987mf}
A.~Schwimmer and N.~Seiberg,
``Comments On The N=2, N=3, N=4 Superconformal Algebras In Two-Dimensions,''
Phys.\ Lett.\ {\bf B184} (1987) 191.
%%CITATION = PHLTA,B184,191;%%

%\cite{Goddard:1988wv}
\bibitem{Goddard:1988wv}
P.~Goddard and A.~Schwimmer,
``Factoring Out Free Fermions And Superconformal Algebras,''
Phys.\ Lett.\ {\bf B214} (1988) 209.
%%CITATION = PHLTA,B214,209;%%

%\cite{Kazama:1989qp}
\bibitem{Kazama:1989qp}
Y.~Kazama and H.~Suzuki,
``New N=2 Superconformal Field Theories And Superstring Compactification,''
Nucl.\ Phys.\ {\bf B321} (1989) 232.
%%CITATION = NUPHA,B321,232;%%

%\cite{Kazama:1989uz}
\bibitem{Kazama:1989uz}
Y.~Kazama and H.~Suzuki,
``Characterization Of N=2 Superconformal Models Generated By Coset Space Method,''
Phys.\ Lett.\ {\bf B216} (1989) 112.
%%CITATION = PHLTA,B216,112;%%


\end{thebibliography}
\end{document}